\documentclass[10pt,journal,compsoc]{IEEEtran}
\ifCLASSOPTIONcompsoc
  \usepackage[nocompress]{cite}
\else
  \usepackage{cite}
\fi
\ifCLASSINFOpdf
\else
\fi

\usepackage{amssymb}
\usepackage{amsmath}
\usepackage{bm}
\usepackage{multirow}
\usepackage{bigstrut}
\usepackage{array}
\usepackage{gensymb}
\usepackage{pifont}
\usepackage{subfigure}
\usepackage{graphicx}
\usepackage{psfrag}
\usepackage{pstricks}
\usepackage{algorithm}
\usepackage{algorithmic}
\usepackage{microtype}
\usepackage[mathscr]{euscript}
\usepackage{enumitem}
\usepackage{sidecap}

\usepackage[bookmarks,backref=true,linkcolor=black]{hyperref}
\hypersetup{
	pdfauthor = {},
	pdftitle = {},
	pdfsubject = {},
	pdfkeywords = {},
	colorlinks=true,
	linkcolor= black,
	citecolor= black,
	pageanchor=true,
	urlcolor = black,
	plainpages = false,
	linktocpage
}

\newcommand{\bfc}{\mathbf{c}}

\newcommand{\bfg}{\mathbf{g}}

\newcommand{\bfn}{\mathbf{n}}

\newcommand{\eigen}{\textsc{Eigen}}
\newcommand{\meanangleerror}{\delta}
\newcommand{\meandist}{\mathscr{D}_{\textrm{mean}}}
\DeclareMathOperator{\diagmat}{diag}
\newcommand{\para}[1]{\textbf{{#1}.~}}

\newcommand{\shortcite}[1]{\cite{#1}}

\begin{document}
%
\title{Static/Dynamic Filtering for Mesh Geometry}
%
%
%
%

\author{Juyong~Zhang,~
        Bailin~Deng$^\dagger$,~
        Yang~Hong,~
        Yue~Peng,~
        Wenjie~Qin,~
        and~Ligang~Liu
\IEEEcompsocitemizethanks{\IEEEcompsocthanksitem J. Zhang, Y. Hong, Y. Peng, W. Qin, and L. Liu are with School of Mathematical Sciences,
	University of Science and Technology of China.\protect\\
\IEEEcompsocthanksitem B. Deng is with  School of Computer Science and Informatics, Cardiff University.}
\thanks{$^\dagger$Corresponding author. Email: \href{mailto:DengB3@cardiff.ac.uk}{\texttt{DengB3@cardiff.ac.uk}}.}
}

%
%

\markboth{~}%
{~}
%



\IEEEtitleabstractindextext{%
\begin{abstract}
The joint bilateral filter, which enables feature-preserving signal smoothing according to the structural information from a guidance, has been applied for various tasks in geometry processing. Existing methods either rely on a static guidance that may be inconsistent with the input and lead to unsatisfactory results, or a dynamic guidance that is automatically updated but sensitive to noises and outliers. Inspired by recent advances in image filtering, we propose a new geometry filtering technique called \emph{static/dynamic filter}, which utilizes both static and dynamic guidances to achieve state-of-the-art results. The proposed filter is based on a nonlinear optimization that enforces smoothness of the signal while preserving variations that correspond to features of certain scales. We develop an efficient iterative solver for the problem, which unifies existing filters that are based on static or dynamic guidances. The filter can be applied to mesh face normals followed by vertex position update, to achieve scale-aware and feature-preserving filtering of mesh geometry. It also works well for other types of signals defined on mesh surfaces, such as texture colors. Extensive experimental results demonstrate the effectiveness of the proposed filter for various geometry processing applications such as mesh denoising, geometry feature enhancement, and texture color filtering.
\end{abstract}

}

\maketitle

\IEEEdisplaynontitleabstractindextext

%
\IEEEpeerreviewmaketitle

\IEEEraisesectionheading{\section{Introduction}}

\IEEEPARstart{S}{ignal} filtering, the process of modifying signals to achieve desirable properties, has become a fundamental tool for different application areas. In image processing, for example, various filters have been developed for smoothing images while preserving sharp edges. Among them, the \emph{bilateral filter}~\cite{bilateral1998} updates an image pixel using the weighted average of nearby pixels, taking into account their spatial and range differences. Its simplicity and effectiveness makes it popular in image processing, and inspires various follow-up work with improved performance~\cite{eisemann2004flash,petschnigg2004digital,cho2014bilateral,ZhangSXJ14}.

\begin{figure*}[!t]
	\includegraphics[width=\textwidth]{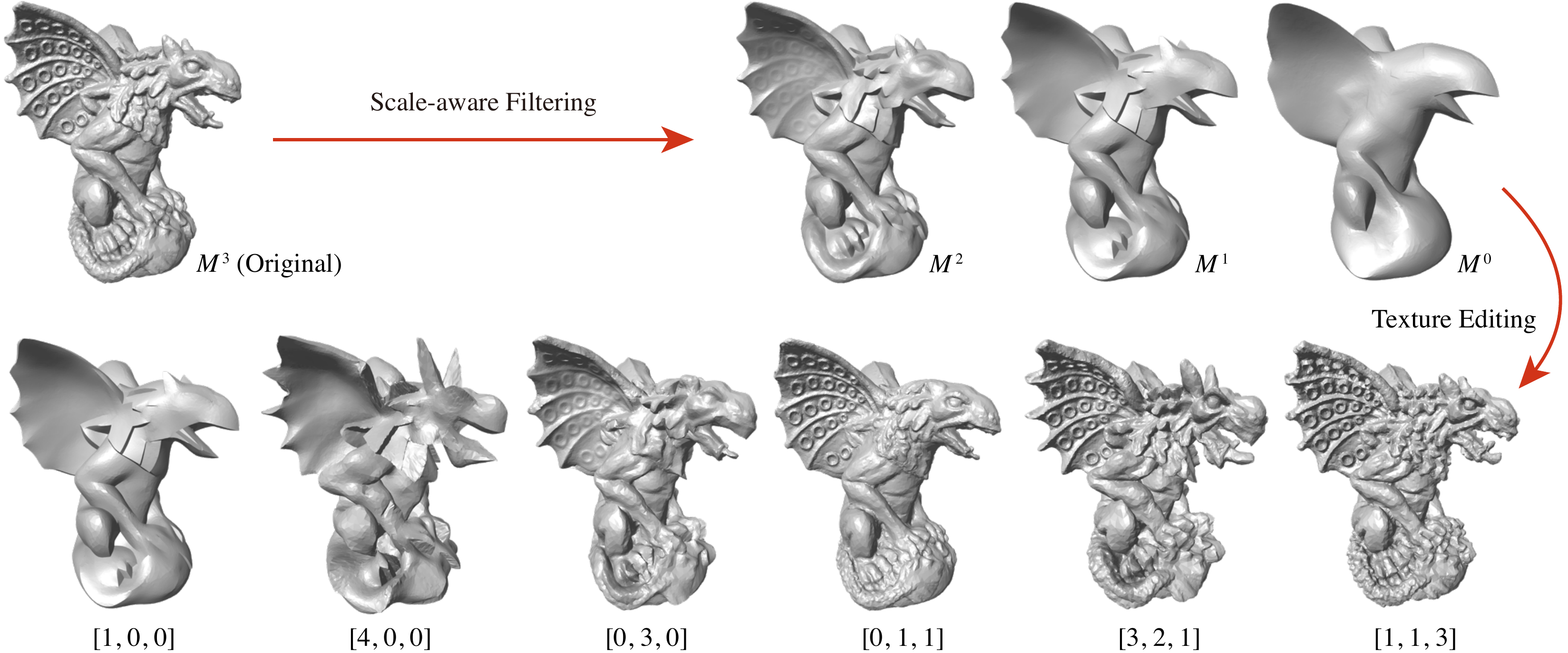}
	\caption{Our SD filter can be used for scale-aware filtering of mesh geometry, allowing us to separate geometry signals according to their scales. Such decomposition can be used for manipulating geometric details, boosting or attenuating features at different scales.}
	\label{fig:teaser}
\end{figure*}

Besides image processing, filtering techniques have also been utilized for processing 3D geometry. Indeed, many geometric descriptors such as normals and vertex positions can be considered as signals defined on two-dimensional manifold surfaces, where image filtering methods can be naturally extended and applied. For example, the bilateral filter has been adapted for feature-preserving mesh smoothing and denoising~\cite{fleishman2003bilateral,jones2003non,zheng2011bilateral,solomon2014general}. 

Development of new geometry filters has also been inspired by other techniques that improve upon the original bilateral filter.
Among them, the \emph{joint bilateral filter}~\cite{eisemann2004flash,petschnigg2004digital} determines the filtering weights using the information from a \emph{guidance} image instead of the input image, and achieves more robust filtering results when the guidance provides reliable structural information. One limitation of this approach is that the guidance image has to be specified beforehand, and remains \emph{static} during the filtering processing. For image texture filtering, Cho et al.~\shortcite{cho2014bilateral} address this issue by computing the guidance using a patch-based approach that reliably captures the image structure. This idea was later adopted by Zhang et al.~\shortcite{ZhangDZBL15} for mesh denoising, where a patch-based guidance is computed for filtering the face normals.  Another improvement for the joint bilateral filter is the \emph{rolling guidance filter} proposed by~\cite{ZhangSXJ14}, which iteratively updates an image using the previous iterate as a \emph{dynamic} guidance, and is able to separate signals at different scales. Recently, this approach was adapted by Wang et al.~\shortcite{WangFLTLG15} to derive a \emph{rolling guidance normal filter} (RGNF), with impressive results for scale-aware geometric processing. 

For guided filtering, the use of static vs dynamic guidance presents a trade-off between their properties. Static guidance enables direct and intuitive control over the filtering process, but is not trivial to construct a priori for general shapes. Dynamic guidance, such as the one used in RGNF, is automatically updated according to the current signal values, but can be less robust when there are outliers or noises in the input signal. Recently, Ham et al.~\shortcite{ham2015} combine static and dynamic guidance for robust image filtering. Inspired by their work, we propose in this paper a new approach for filtering signals defined on mesh surfaces, by utilizing both static and dynamic guidances. The filtered signal is computed by minimizing a target function that enforces consistency of signal values within each neighborhood, while incorporating structural information provided by a static guidance. To solve the resulting noncovex optimization problem, we develop an efficient fixed-point iteration solver, which significantly outperforms the majorization-minimization (MM) algorithm proposed by~\cite{ham2015} for similar problems. Moreover, unlike the MM algorithm, our solver can handle constraints such as unit length for face normals, which are important for geometry processing problems. 
Our solver iteratively updates the signal values by combining the original signal with the current signal from a spatial neighborhood. The combination weights are determined according to the static input guidance as well as a dynamic guidance derived from the current signal. The proposed method, called \emph{static/dynamic} (SD) filtering, benefits from both types of guidance and produces scale-aware and feature-preserving results.

The proposed method can be applied to different signals on mesh surfaces. When applied to face normals followed by vertex updates, it filters geometric features according to their scales. When applied to mesh colors obtained from texture mapping, it filters the colors based on the metric on the mesh surface. In addition, utilizing the scale-awareness of the filter, we apply it repeatedly to separate signal components of different scales; the results can be combined according to user-specified weights, allowing for intuitive feature manipulation and enhancement. Extensive experimental results demonstrate the efficiency and effectiveness of our filter. We also release the source codes to ensure reproducibility.

In addition, we propose a new method for vertex update according to face normals, using a nonlinear optimization formulation that enforces the face normal conditions while preserving local triangle shapes. The vertex positions are computed by iteratively solving a linear system with a fixed sparse positive definite matrix, which is done efficiently via pre-factorization of the matrix. Compared with existing approaches, our method produces meshes that are more consistent with the filtered face normals. 

In summary, our main contributions include:
\begin{itemize}[leftmargin=*]
	\item we extend the work of Ham et al.~\shortcite{ham2015} and propose an SD filter for signals defined on triangular meshes, formulated as an optimization problem;
	\item we develop an efficient fixed-point iteration solver for the SD filter, which can handle constraints such as unit normals and significantly outperforms the MM solver from~\cite{ham2015};
	\item we propose an efficient approach for updating vertex positions according to filtered face normals, which produces new meshes that are consistent with the target normals while preserving local triangle shapes;
	\item based on the SD filter, we develop a method to separate and combine signal components of different scales, enabling intuitive feature manipulation for mesh geometry and texture color.
\end{itemize}

\section{Related Work}

In the past, various filtering approaches have been proposed to process mesh geometry. Early work from Taubin~\shortcite{Taubin1995} and Desbrun et al.~\shortcite{Desbrun1999} applied low-pass filters on meshes, which remove high-frequency noises but also attenuate sharp features. 
Later, Taubin~\shortcite{Taubin2001} proposed a two-step approach that first performs smoothing on face normals, followed by vertex position updates using anisotropic filters.
To enhance crease edges, Ohtake et al.~\shortcite{Ohtake2001} applied anisotropic diffusion to mesh normals before updating vertex positions.
Chuang and Kazhdan~\cite{ChuangK11} developed a framework for curvature-aware mesh filtering based on the screened Poisson equation.

An important class of mesh filtering techniques is based on the bilateral filter~\cite{bilateral1998}. On images, the bilateral filter updates a pixel using a weighted average of its neighboring pixels, with larger contribution from pixels that are closer in spatial or range domain. It can smooth images while preserving edges where there is large difference between neighboring pixel values~\cite{paris2009bilateral}. Different methods have been developed to adapt the bilateral filter to mesh geometry. Fleishman et al.~\shortcite{fleishman2003bilateral} and Jones et al.~\shortcite{jones2003non} applied the bilateral filter to the mesh vertex positions for feature-preserving mesh denoising. Zheng et al.~\shortcite{zheng2011bilateral} applied the bilateral filter to mesh face normals instead, followed by vertex position update to reconstruct the mesh shape. Solomon et al.~\shortcite{solomon2014general} proposed a framework for bilateral filter that is applicable for signals on general domains including images and meshes, with a rigorous theoretical foundation. Besides denoising, bilateral filtering has also been applied for other geometry processing applications such as point cloud normal enhancement~\cite{Jones2004} and mesh feature recovery~\cite{Wang2006}.

The bilateral filter inspired a large amount of follow-up work on image filtering. Among them, the joint bilateral filter~\cite{eisemann2004flash,petschnigg2004digital} extends the original bilateral filter by evaluating the spatial kernel using a guidance image. It can produce more reliable results when the guidance image correctly captures the structural information of the target signal. This property was utilized by Eisemann \& Durand~\shortcite{eisemann2004flash} and Petschnigg et al.~\cite{petschnigg2004digital} to filter flash photos, using corresponding non-flash photos as the guidance. Kopf et al.~\shortcite{KopfCLU07} and Cho et al.~\shortcite{cho2014bilateral} applied the joint bilateral filter for image upsampling and structure-preserving image decomposition, respectively.  In particular, a patch-based guidance is constructed in~\shortcite{cho2014bilateral} to capture the input image structure. This idea was later adopted in~\cite{ZhangDZBL15} for filtering mesh face normals, where the guidance normals are computed using surface patches with the most consistent normals.  Zhang et al.~\shortcite{ZhangSXJ14} proposed a different approach to guidance construction in their iterative \emph{rolling guidance filter}, where the resulting image from an iteration is used as a dynamic guidance for the next iteration. The rolling guidance filter produces impressive results for scale-aware image processing, and is able to filter out features according to their scales. Wang et al.~\cite{WangFLTLG15} adapted this approach to filter mesh face normals; the resulting \emph{rolling guidance normal filter} enables scale-aware processing of geometric features, but is sensitive to noises on the input model. Recently, Ham et al.~\shortcite{ham2015} proposed a robust image filtering technique based on an optimization formulation that involves a nonconvex regularizer. Their technique is effectively an iterative filter that incorporates both static and dynamic guidances, and achieves superior results in terms of robustness, feature-preservation, and scale-awareness. Our SD filter is based on a similar optimization formulation, but takes into account the larger filtering neighborhoods that are necessary for geometry signals. It enjoys the same desirable properties as its counterpart in image processing. In addition, the numerical solver proposed in~\cite{ham2015} can only handle unconstrained signals, and is less efficient for the large neighborhoods used in our formulation. We therefore propose a new solver that outperforms the one from~\cite{ham2015}, while allowing for constrained signals such as unit normals. 

Feature-preserving signal smoothing can also be achieved via optimization. Notable examples include image smoothing algorithms that induce sparsity of image gradients via $\ell_0$-norm~\cite{xu2011image} or $\ell_1$-norm~\cite{RUDIN1992} regularization. These approaches were later adapted for mesh smoothing and denoising~\cite{Taubin12c,he2013mesh,Zhang2015Variational}. Although effective in many cases, their optimization formulation only regularizes the signal difference between immediately neighboring faces. In comparison, our optimization compares signals within a neighborhood with user-specified size, which provides more flexibility and achieves better preservation of large-scale features.

From a signal processing point of view, meshes can be seen as a combination of signals with multiple frequency bands, which also relates with the scale space analysis~\cite{Perona90}. Previous work separate geometry signals of different frequencies using eigenfunctions of the heat kernel~\cite{sun2009concise} or the Laplace operator~\cite{Vallet2008,zhang2010spectral}. Although developed with sound theoretical foundations, such approaches are computationally expensive. Moreover, as specific geometric features can span across a wide range of frequencies, it is not easy to preserve or manipulate them with such approaches. The recent work from Wang et al.~\cite{WangFLTLG15} provides an efficient way to separate and edit geometric features of different scales, harnessing the scale-aware property of the rolling guidance filter. Our SD filter also supports  scale-aware processing of geometry signals, with more robustness than RGNF thanks to the incorporation of both static and dynamic guidances.

\section{The SD Filter}
\label{sec:filter}   
\begin{figure}[!h]
	\centering
	\includegraphics[width=0.7\columnwidth]{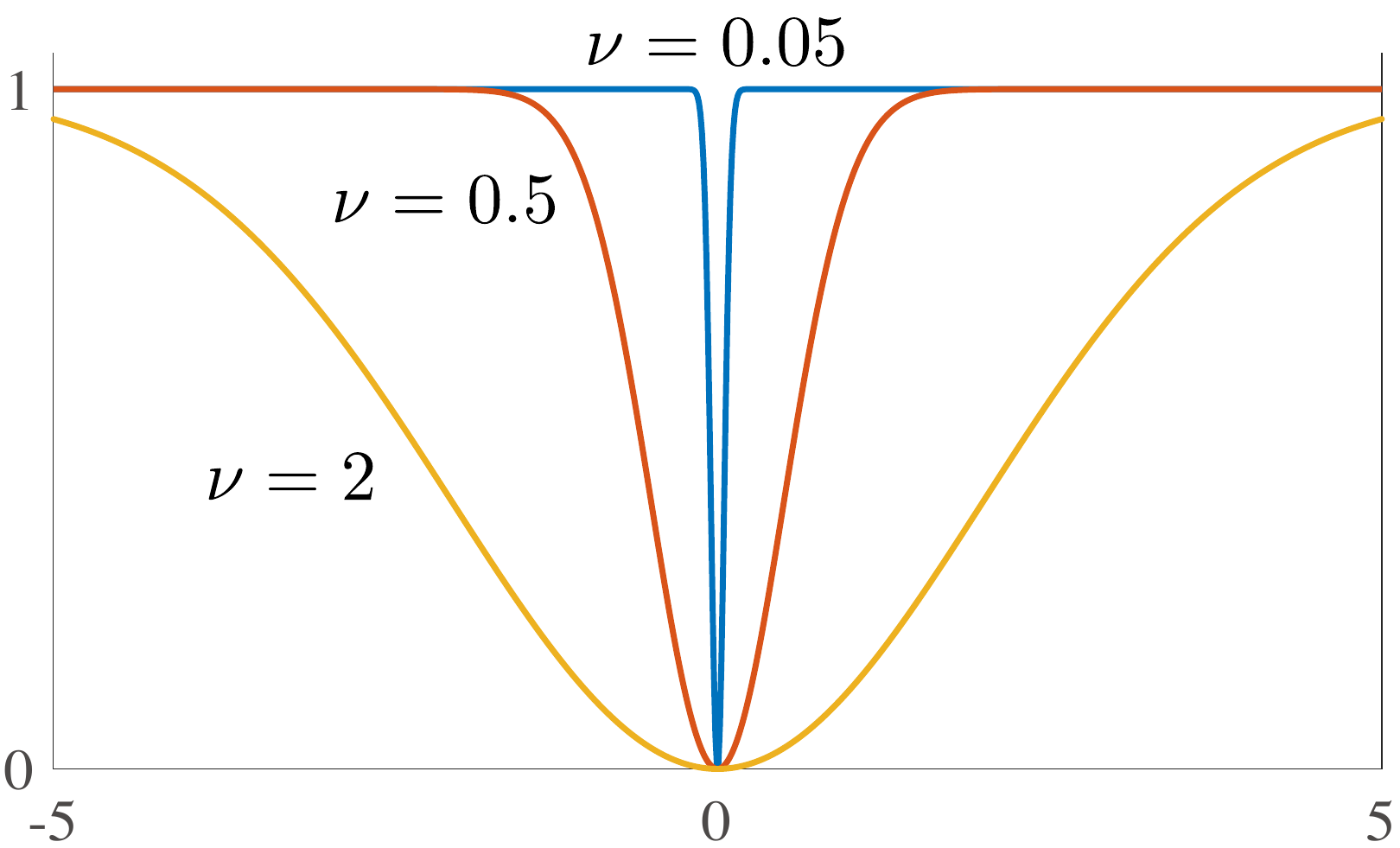}
	\caption{The graphs of $\psi_{\nu}(x)$ with different $\nu$ parameters.}
	\label{fig:RegularizationFunction}
\end{figure}

The SD filter was originally proposed by Ham et al.~\shortcite{ham2015} for robust image processing. Given an input image $F$ and a static guidance image $G$, they compute an output image $U$ via  optimization
\begin{equation}
\min_U~\sum_i \gamma_i (U_i - F_i)^2 + ~\lambda \sum_{(i,j) \in \mathcal{N}} \phi_{\mu}(G_i - G_j) \cdot \psi_{\nu}(U_i - U_j),
\label{eq:SDImageFilter}
\end{equation}
where $F_i, G_i, U_i$ are the pixel values of $F$, $G$ and $U$ respectively, $\gamma_i$ and $\lambda$ are user-specified weights, $\mathcal{N}$ denotes the set of 8-connected neighboring pixels, and
\begin{equation}
\phi_{\mu}(x) = \exp(-\displaystyle\frac{x^{2}}{2\mu^{2}}), \quad \psi_{\nu}(x) = 1 - \phi_{\nu}(x). 
\end{equation}
The first term in the target function is a fidelity term that requires the output image to be close to the input image, while the second term is a regularizer for the output image. Function $\psi_{\nu}$ (see Fig.~\ref{fig:RegularizationFunction}) penalizes the difference between adjacent pixels, but with bounded penalty for pixel pairs with large difference which correspond to edges or outliers. When $\nu$ approaches $0$, $\psi_{\nu}$  approaches the $\ell_0$ norm. Function $\phi_{\mu}$ is a Gaussian weight according to the guidance, with larger weights for pixel pairs with closer guidances. Thus the regularizer promotes smooth regions and preserves sharp features based on the guidance, and is robust to outliers.

In this paper, we propose an SD filter for signals defined on 2-manifold surfaces represented as triangular meshes. We begin our discussion with filtering face normals, a common approach for smoothing mesh geometry~\cite{sun2007fast,zheng2011bilateral,solomon2014general,ZhangDZBL15}.

\subsection{SD filter for face normals}
For a given orientable triangular mesh, let $\mathbf{n}_i \in \mathbb{R}^3$ be the outward unit normal of face $f_i$, computed as
\begin{equation}
\mathbf{n}_i = \frac{(\mathbf{v}_{i_1} - \mathbf{v}_{i_2}) \times (\mathbf{v}_{i_3} - \mathbf{v}_{i_2})}{\|(\mathbf{v}_{i_1} - \mathbf{v}_{i_2}) \times (\mathbf{v}_{i_3} - \mathbf{v}_{i_2})\|},
\label{eq:OrientedNormal}
\end{equation}
where $\mathbf{v}_{i_1}, \mathbf{v}_{i_2}, \mathbf{v}_{i_3} \in \mathbb{R}^3$ are its vertex positions enumerated according to the orientation. We associate the normal with the face centroid $\mathbf{c}_i = \frac{1}{3}\sum_{k=1}^3 \mathbf{v}_{i_k}$. We first filter the face normals, and then update the mesh vertices accordingly. To define an SD filter for the normals $\{\mathbf{n}_i\}$, we must consider some major differences compared with image filtering:
\begin{itemize}[leftmargin=*]
	\item Image pixels are located on a regular grid, but mesh faces may result from irregular sampling of the surface.
	\item To smooth an image, the SD filter as per Eq.~\eqref{eq:SDImageFilter} only considers the difference between a pixel and its eight neighbor pixels. On meshes, however, geometry features can span across a large region, thus we may need to compare face normals beyond one-ring neighborhoods~\cite{WangFLTLG15}. Moreover, similar to the bilateral filter, such comparison should consider the difference between the spatial locations, with stronger penalty for normal deviation between faces that are closer to each other.
\end{itemize}
Therefore, we compute the filtered normals $\{\mathbf{n}_i\}$ by minimizing a target function 
\begin{equation}
E_{\textrm{SD}} =  E_{\textrm{fid}} + \lambda E_{\textrm{reg}},
\label{eq:objective_energy}
\end{equation}
with a user-specified weight $\lambda > 0$.
Here $E_{\textrm{fid}}$ is a fidelity term between the input and output normals,
\begin{equation}
E_{\textrm{fid}} = \sum_{i} A_i \| \mathbf{n}_i - \hat{\mathbf{n}}_i \|^2,
\end{equation}
where $\hat{\mathbf{n}}_i, A_i$ are the normal and area of face $f_i$, respectively. $E_{\textrm{reg}}$ is a regularization term defined as
\begin{equation}
\begin{split}
E_{\textrm{reg}} = \sum_i \sum_{f_j \in N(i)} & [~A_j \cdot \phi_{\eta}(\|\mathbf{c}_i - \mathbf{c}_j\|)\\
& ~~~~ \cdot \phi_{\mu}(\|\mathbf{g}_i - \mathbf{g}_j\|) \cdot \psi_{\nu}(\|\mathbf{n}_i - \mathbf{n}_j\|)~],
\end{split}
\end{equation}
where $\{\mathbf{g}_i\}$ are the guidance face normals, and $N(i)$ denotes the set of neighboring faces of $f_i$. The Gaussian standard deviation parameters $\eta, \mu, \nu \in \mathbb{R}^{+}$ are controlled by the user. Compared with the image regularizer in Eq.~\eqref{eq:SDImageFilter}, this formulation introduces a Gaussian weight $\phi_{\eta}$ for the spatial locations of face normals. Here $\phi_{\eta}$ is defined according to the Euclidean distance between face centroids for  simplicity of computation, but other distance measures such as the geodesic distance can also be used. For each face $f_i$, its neighborhood $N(i)$ is chosen to be the set of faces with a significant value of the spatial weight $\phi_{\eta}(\|\mathbf{c}_i - \mathbf{c}_j\|)$. Using the empirical three-sigma rule~\cite{Pukelsheim1994}, we include in $N(i)$ the faces $\{f_j\}$ with $\|\mathbf{c}_j - \mathbf{c}_i\| \leq 3 \eta$, which can be found using a breadth-first search from $f_i$. 

The target function $E_{\textrm{SD}}$ is nonconvex because of $\psi_{\nu}$, and needs to be minimized numerically. In the following, we first show how the majorization-minimization (MM) algorithm proposed in~\cite{ham2015} can be extended to solve this problem. Afterwards, we propose a new fixed-point iteration solver that significantly outperforms the MM algorithm and is suitable for interactive applications.

\para{MM algorithm}
For the SD image filter, Ham et al.~\shortcite{ham2015} proposed a majorization-minimization (MM) algorithm to iteratively minimize the target function~\eqref{eq:SDImageFilter}. In each iteration, the target function is replaced by a convex surrogate function that bounds it from above, which is computed using the current variable values. This surrogate function is then minimized to update the variables. The MM solver is guaranteed to converge to a local minimum of the target function.
Thus a straightforward way to minimize the new target function~\eqref{eq:objective_energy} is to employ the MM algorithm, using the convex surrogate function $\Psi^{t}_{\nu}$ for $\psi_{\nu}(x)$ at $x = t$~\cite{ham2015}:
\begin{equation}
\Psi^{t}_{\nu}(x) = \psi_{\nu}(t) + (x^2 - t^2)(1/2\nu^2 - \psi_{\nu}(t)/2\nu^2).
\label{eq:Surrogate}
\end{equation}
Specifically, with the variable values $\left\{\mathbf{n}_i^{k}\right\}$ at iteration $k$, we replace the term $\psi_{\nu}(\|\mathbf{n}_i - \mathbf{n}_j\|)$ in the target function by its convex surrogate $\Psi^{\|\mathbf{n}^k_i - \mathbf{n}^k_j\|}_{\nu}(\|\mathbf{n}_i - \mathbf{n}_j\|)$ according to Eq.~\eqref{eq:Surrogate}. The updated variable values $\left\{\mathbf{n}_i^{k+1}\right\}$ are computed from the resulting convex problem
\begin{equation}
\min_{\{\mathbf{n}_i^{k+1}\}}~~\sum_i A_i \|\mathbf{n}_i^{k+1} - \hat{\mathbf{n}}_i\|^2  + \lambda \sum_i \sum_{f_j\in N(i)} w_{ij}^k \|\mathbf{n}_i^{k+1} - \mathbf{n}_j^{k+1}\|_{2}^2,
\label{eq:MMTarget}
\end{equation}
where 
\begin{equation}
w^k_{ij} =  \frac{A_j}{2\nu^2} \cdot \phi_{\eta}(\|\bfc_i - \bfc_j\|)\cdot \phi_{\mu}(\|\bfg_i - \bfg_j\|) \cdot \phi_{\nu}(\|\mathbf{n}_i^k - \mathbf{n}_j^k\|).
\label{eq:weghts}
\end{equation}
Due to the symmetry of neighboring relation between faces (i.e., $f_j \in N(i) \Leftrightarrow f_i \in N(j)$), the optimization problem~\eqref{eq:MMTarget} amounts to solving a linear system:
\begin{equation}
(\mathbf{D} + \lambda\mathbf{M}^k) \mathbf{N}^{k+1} = \mathbf{D} \hat{\mathbf{N}},
\label{eq:LinearSystem}
\end{equation}
where $\mathbf{D} = \diagmat(A_1, A_2, \ldots, A_{n_f})$ with $n_f$ being the number of faces, $\mathbf{N}^{k+1}, \hat{\mathbf{N}} \in \mathbb{R}^{n_f \times 3}$ stack the values of $\{\mathbf{n}_i^{k+1}\}$ and $\{\hat{\mathbf{n}}_i\}$ respectively, and $\mathbf{M}^k \in \mathbb{R}^{n_f \times n_f}$ is a symmetric matrix with diagonal elements 
\[
m_{ii}^k = \sum_{j \in N(i)} \left(w_{ij}^k + w_{ji}^k\right),
\]
and off-diagonal elements
\[
m_{ij}^k
= 
\left\{
\begin{array}{ll}
- w_{ij}^k - w_{ji}^k, & \textrm{if}~ j \in N(i),\\
0, & \textrm{otherwise}.
\end{array}
\right.
\]
The linear system matrix in Eq.~\eqref{eq:LinearSystem} is diagonally dominant and symmetric positive definite, and can be solved using standard linear algebra routines. 

\para{Fixed-point iteration solver} Although the MM algorithm works well on images, its performance on meshes is often unsatisfactory. Due to larger face neighborhoods, there are a large number of nonzeros in the linear system matrix of Eq.~\eqref{eq:LinearSystem}, resulting in long computation time for each iteration. In the following, we propose a more efficient solver that is suitable for interactive applications. Note that a local minimum of the target function~\eqref{eq:objective_energy} should satisfy the first order optimality condition $\partial E_{\textrm{SD}} / \partial \mathbf{n}_i = \mathbf{0}$ for each $\mathbf{n}_i$, which expands into
\begin{equation}
A_i (\mathbf{n}_i  - \hat{\mathbf{n}}_i) +  \lambda \sum_{f_j \in N(i)} b_{ij} (\mathbf{n}_i - \mathbf{n}_j) = \mathbf{0}, \quad \forall~i, 
\label{eq:OptimalityCondition}
\end{equation}
where
\begin{equation}
b_{ij} = \frac{A_i + A_j}{2 \nu^2} \phi_{\eta}(\|\bfc_i - \bfc_j\|) \cdot \phi_{\mu }(\| \bfg_i - \bfg_j\|) \cdot \phi_{\nu}(\|\bfn_i - \bfn_j\|).
\end{equation}
The equations~\eqref{eq:OptimalityCondition} can be solved using fixed-point iteration
\begin{equation}
\bfn_i^{k+1} = \frac{A_i \hat{\bfn}_i + \lambda \sum_{f_j\in N(i)}  b^k_{ij} \bfn^k_{j}}{A_i + \lambda \sum_{f_j\in N(i)} b^k_{ij}},
\label{eq:FixedPointIterationFormat}
\end{equation}
with $b_{ij}^k = \frac{A_i + A_j}{2 \nu^2} \phi_{\eta}(\|\bfc_i - \bfc_j\|) \cdot \phi_{\mu }(\| \bfg_i - \bfg_j\|) \cdot \phi_{\nu}(\|\bfn_i^k - \bfn_j^k\|)$.
In this way, the updated normal $\bfn_i^{k+1}$ of a face $f_i$ is a convex combination of its initial normal $\hat{\bfn}_i$ and the current normals $\{\bfn^k_{j}\}$ of the faces in its neighborhood. The convex combination coefficient for a neighboring face normal $\bfn^k_{j}$ depends on both the (static) difference between the guidance normals $(\bfg_i, \bfg_j)$ on the two faces, and the (dynamic) difference between their current normals $(\bfn^k_{i}, \bfn^k_{j})$, hence the name \emph{static/dynamic} filter. Moreover, there is an interesting connection between the fixed-point iteration and the MM algorithm: the iteration~\eqref{eq:FixedPointIterationFormat} is a single step of Jacobi iteration for solving the MM linear system~\eqref{eq:LinearSystem}.

\begin{figure}[!t]
	\centering
	\includegraphics[height=0.68\columnwidth]{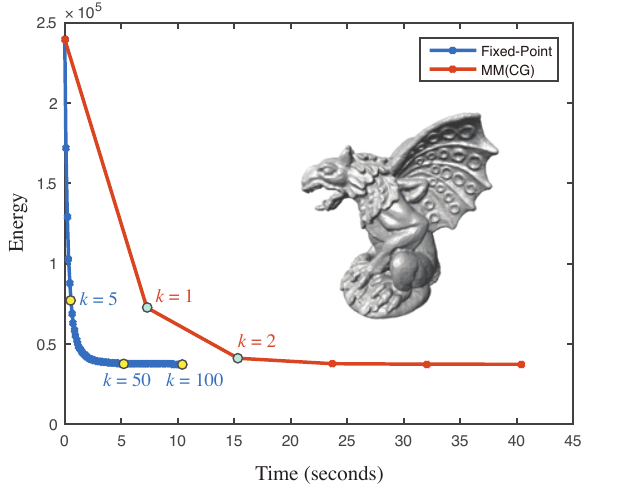}
	\caption{The change of target energy $E_{\textrm{SD}}$ for the Gargoyle model with respect to the computational time, using the fixed-point iteration solver~\eqref{eq:FixedPointIterationFormat} for 100 iterations and the MM algorithm for 5 iterations, respectively. The fixed-point iteration solver takes much shorter time per iteration, and reduces the energy to a value close to the solution within a fraction of the time for one MM iteration.}
	\label{fig:CompareWithMM}
\end{figure}

It can be shown that the fixed-point iteration monotonically decreases the target function value until it converges to a local minimum. A proof is provided in the supplementary material. Moreover, the face normal updates are trivially parallelizable, enabling speedup on GPUs and multi-core CPUs. Our fixed-point solver is significantly faster than the MM algorithm, as shown in Fig.~\ref{fig:CompareWithMM} where we compare the change of target energy with respect to the computational time between the two solvers. Here the spatial Gaussian parameter $\eta$ is set to three times the average distance between neighboring face centroids in the mesh. To achieve the best performance for the MM solver, we tested two strategies for solving the MM linear system: 1) pre-computing symbolic Cholesky factorization of the system matrix, and performing numerical factorization in each iteration, with the three right-hand-sides for $x$-, and $y$- and $z$-coordinates solved in parallel; 2) conjugate gradient method with parallel sparse matrix-vector multiplication. Due to the large neighborhood size, the MM system matrix has a large number of non-zeros in each row, and the Cholesky factorization approach is much more time-consuming than conjugate gradient. Therefore, we only show the conjugate gradient timing. We can see that the fixed-point solver is much more efficient than the MM algorithm, drastically reducing the energy to a value close to the solution within a fraction of the computational time for one MM iteration. This phenomenon is observed in our experiments with other models as well. Detailed results are provided in the supplemental materials.

\begin{figure}[!t]
	\centering
	\includegraphics[height=0.68\columnwidth]{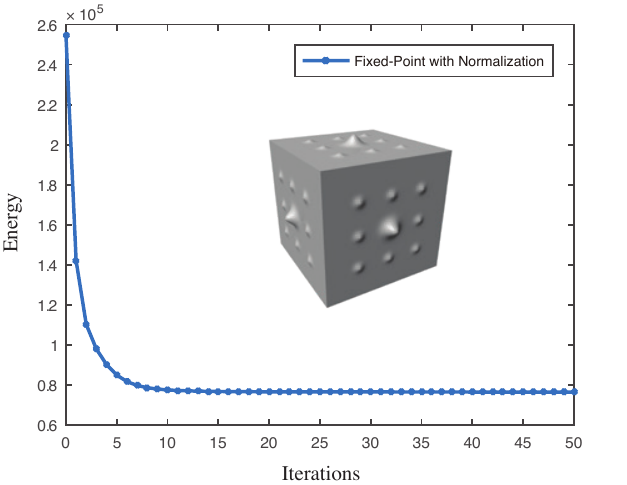}
	\caption{The change of the modified target energy $\overline{E}_{\textrm{SD}}$ with respect to the number of iterations, using our fixed-point iteration solver~\eqref{eq:NormalizedFixedPoint} for the cube model in Fig.~\ref{fig:TextureRemoval-Cube}. The solver rapidly decreases the energy within a small number of iterations.}
	\label{fig:IterEnergy}
\end{figure}

\begin{figure*}[!t]
	\centering
	\includegraphics[width=\textwidth]{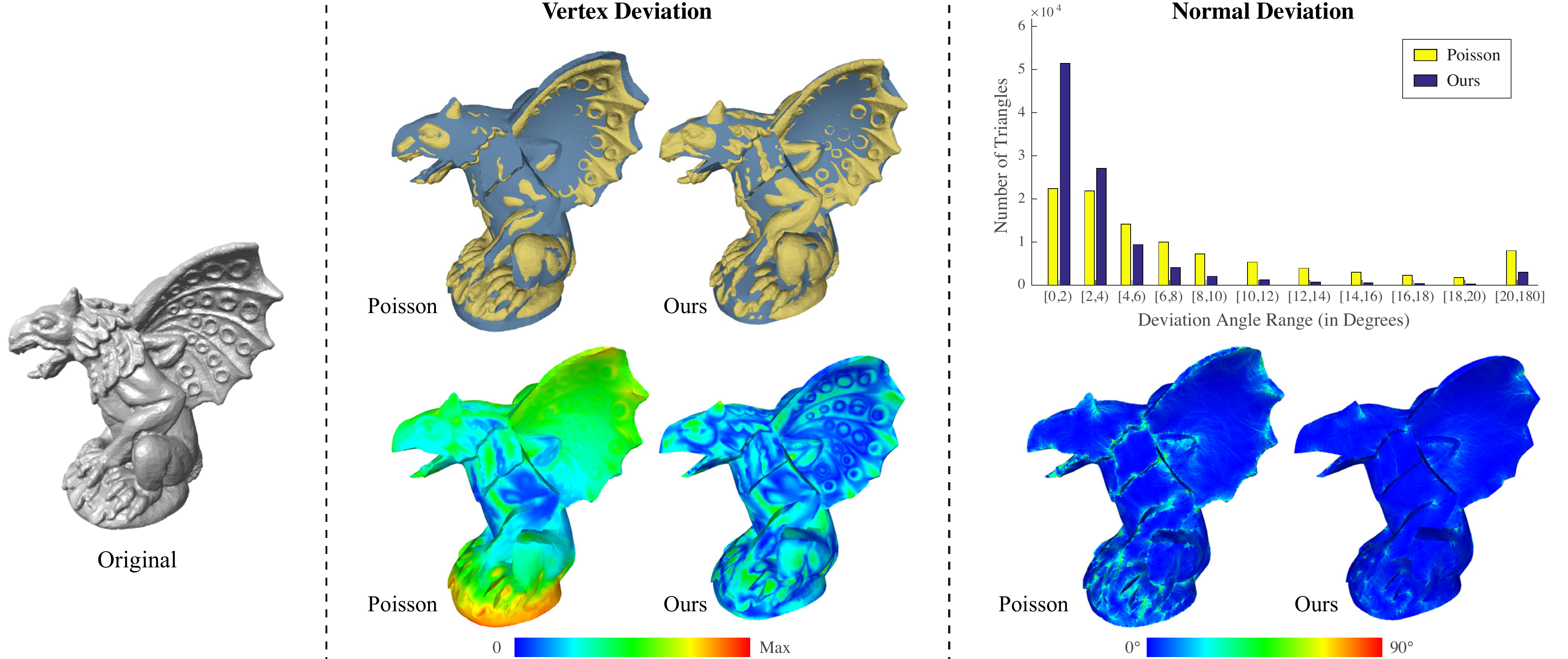}
	\caption{Comparison between our vertex update method and the Poisson-based update method from~\protect\cite{WangFLTLG15}, based on the same original mesh (left) and target normals. Top center: we align the centroid of each resulting mesh (in blue) with the centroid of the original mesh (in yellow) to minimize the $\ell_2$ norm of the deviation between their vertices; the Poisson-based method leads to larger deviation from the original mesh. Bottom center: the deviation between individual vertices after the centroid alignment is visualized via color coding. Right: we evaluate the deviation between the resulting normals and target normals for each face, and visualize its distribution across the mesh via histogram (top right) and color coding (bottom right).}
	\label{fig:VertexUpdateComparison}
\end{figure*}

\para{Enforcing unit normal constraints} The target function $E_{\textrm{SD}}$ in Eq.~\eqref{eq:objective_energy} adapts the SD filter to mesh face normals in a straightforward way, but fails to recognize the requirement that all normals should lie on the unit sphere. In fact, starting from the unit face normals of the input mesh, $E_{\textrm{SD}}$ can be decreased by simply shrinking the normals without changing their directions, and the filtered normals can have different lengths across the mesh. In other words, without the requirement of unit normals, the difference between two normal vectors is not a reliable measure of the deviation between their directions, which makes $E_{\textrm{SD}}$ less effective for controlling the filter. To resolve this issue, we derive a new target function $\overline{E}_{\textrm{SD}}$ for optimization, by substituting each normal vector $\mathbf{n}_i (i=1,\ldots,n_f)$ in $E_{\textrm{SD}}$ with its normalization $\overline{\mathbf{n}}_i = \mathbf{n}_i / \|\mathbf{n}_i\|$.
However, such normalization increases the nonlinearity of the problem and makes its numerical solution more challenging. The MM algorithm is no longer applicable, because the quadratic surrogate in Eq.~\eqref{eq:Surrogate} does not hold here. On the other hand, the fixed-point iteration solver can be slightly modified to minimize $\overline{E}_{\textrm{SD}}$ efficiently. At a local minimum, the first-order optimality condition $\partial \overline{E}_{\textrm{SD}} / \partial \mathbf{n}_i = \mathbf{0}$ amounts to
\begin{equation}
\left(\mathbf{I}_3 - \overline{\mathbf{n}}_i \overline{\mathbf{n}}_i^T\right)~\left(
\overline{\bfn}_i - \frac{A_i \hat{\bfn}_i + \lambda \sum_{f_j\in N(i)}  \overline{b}_{ij} \overline{\bfn}_{j}}{A_i + \lambda \sum_{f_j\in N(i)} \overline{b}_{ij}} \right) = \mathbf{0}, 
\label{eq:NormalizedOptimalityCondition}
\end{equation}
where
\[
\overline{b}_{ij} = \frac{A_i + A_j}{2 \nu^2} \phi_{\eta}(\|\bfc_i - \bfc_j\|) \cdot \phi_{\mu }(\| \bfg_i - \bfg_j\|) \cdot \phi_{\nu}(\|\overline{\bfn}_i - \overline{\bfn}_j\|), 
\]
and $\mathbf{I}_3$ is the $3 \times 3$ identity matrix. The matrix $\mathbf{I}_3 - \overline{\mathbf{n}}_i \overline{\mathbf{n}}_i^T$ represents the projection onto the subspace orthogonal to $\overline{\mathbf{n}}_i$. Geometrically, condition~\eqref{eq:NormalizedOptimalityCondition} means that the linear combination
$A_i \hat{\bfn}_i + \lambda \sum_{f_j\in N(i)}  \overline{b}_{ij} \overline{\bfn}_{j}$ must be parallel to $\overline{\mathbf{n}}_i$. Therefore, we can update $\mathbf{n}_i$ via
\begin{equation}
\mathbf{n}_i^{k+1} = \frac{A_i \hat{\bfn}_i + \lambda \sum_{f_j\in N(i)}  \overline{b}_{ij}^k \overline{\bfn}_{j}^k}{\|A_i \hat{\bfn}_i + \lambda \sum_{f_j\in N(i)}  \overline{b}_{ij}^k \overline{\bfn}_{j}^k\|},
\label{eq:NormalizedFixedPoint}
\end{equation}
with  $\overline{b}_{ij}^k = \frac{A_i + A_j}{2 \nu^2} \phi_{\eta}(\|\bfc_i - \bfc_j\|) \cdot \phi_{\mu}(\| \bfg_i - \bfg_j\|) \cdot \phi_{\nu}(\|\overline{\bfn}_i^k - \overline{\bfn}_j^k\|)$. Compared with the previous iteration format~\eqref{eq:FixedPointIterationFormat}, this simply adds a normalization step after each iteration.

Similar to the previous fixed-point iteration format, the new solver with Eq.~\eqref{eq:NormalizedFixedPoint} is embarrassingly parallel, and rapidly decreases the target function within a small number of iterations (see Fig.~\ref{fig:IterEnergy}). In the following, all examples of SD normal filtering are processed using this solver.

\para{Vertex update}
After filtering the face normals, we need to update the mesh vertices accordingly. Many existing methods do so by enforcing orthogonality between the new edge vectors and the target face normals~\cite{sun2007fast}. Although very efficient, such methods can result in a large number of flipped triangles, because the orthogonality constraint is still satisfied if the updated face normal is opposite to the target one. To address this issue, we propose a new update method that directly enforces the oriented normals as soft constraints, in the same way as~\cite{Bouaziz2012}. Specifically, for each face $f_i$ with a target oriented unit normal $\hat{\mathbf{n}}_i$, we define $\mathcal{C}_i$ as the feasible set of its vertex positions for which the resulting oriented unit normal is $\hat{\mathbf{n}}_i$. The new vertex positions $\mathbf{V} = [\mathbf{v}_1, \ldots, \mathbf{v}_{n_v}]^T \in \mathbb{R}^{n_v \times 3}$ are determined by solving
\begin{equation}
\min_{\mathbf{V}, \mathbf{P}} ~~ w~\|\mathbf{V} - \mathbf{V}^0\|_F^2 + \sum_{i = 1}^{n_f} \|\mathbf{M} \mathbf{V}_{f_i} - \mathbf{P}_i\|_F^2 + \sigma_i(\mathbf{P}_i).
\label{eq:VertexUpdateTargetFunction}
\end{equation}
Here the first term penalizes the deviation between the new vertex positions and the original vertex positions $\mathbf{V}^0$, with $\|\cdot\|_F$ being the Frobenius norm. $w$ is a user-specified positive weight, which is set to 0.001 by default. Matrix $\mathbf{V}_{f_i} \in \mathbb{R}^{3 \times 3}$ stores the vertex positions of face $f_i$ in its rows. Matrix
\[
\mathbf{M} = 
\frac{1}{3}
\begin{bmatrix}
2 & -1 & -1\\
-1 & 2 & -1\\
-1 & -1 &2
\end{bmatrix}
\]  
produces the mean-centered vertex positions for a face. $\mathbf{P}_i \in \mathbb{R}^{3 \times 3}$ are auxiliary variables representing the closest projection of $\mathbf{M} \mathbf{V}_{f_i}$ onto the feasible set $\mathcal{C}_i$, and $\sigma_i$ is an indicator function for $\mathcal{C}_i$, so that
\[
\sigma_i(\mathbf{P}_i)
= \left\{
\begin{array}{ll}
0 & \textrm{if}~\mathbf{P}_i \in \mathcal{C}_i,\\
+ \infty & \textrm{otherwise}.
\end{array}
\right.
\]
The second term of the target function~\eqref{eq:VertexUpdateTargetFunction} penalizes the violation of oriented normal constraint for each face, using the squared Euclidean distance to the feasible sets. The use of mean-centering matrix $\mathbf{M}$ utilizes the translation-invariance of the oriented normal constraint, to allow for faster convergence of the solver~\cite{Bouaziz2012}. Overall, this optimization problem searches for new vertex positions that satisfy the oriented normal constraints as much as possible, while being close the original positions.
It is solved via alternating minimization of $\mathbf{V}$ and $\mathbf{P}$, following the approach of~\cite{Bouaziz2012}:
\begin{figure*}[!t]
	\centering
	\includegraphics[width=\textwidth]{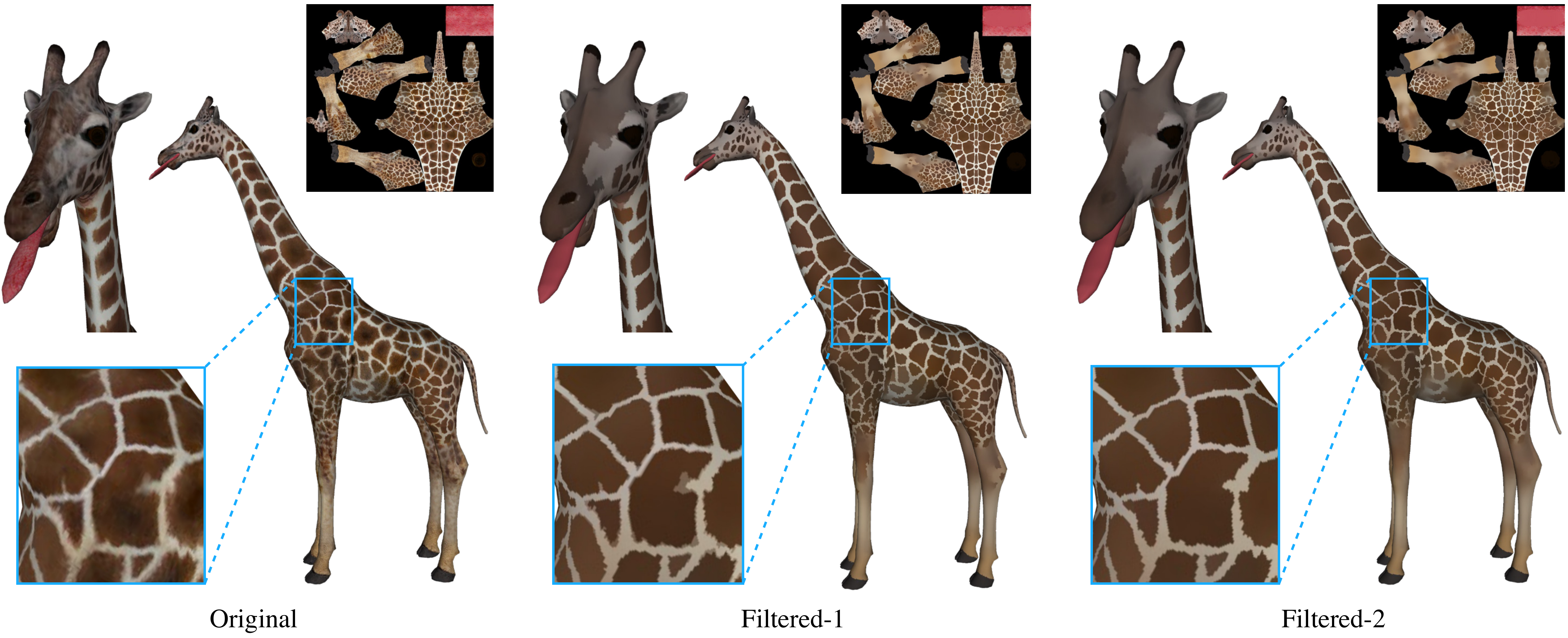}
	\caption{SD filtering of texture image, which can smooth out features based on their scales on the mesh surface.}
	\label{fig:TextureRemove-Giraffe}
\end{figure*}
\begin{itemize}[leftmargin=*]
	\item First, we fix $\mathbf{V}$ and update $\mathbf{P}$. This reduces to a set of separable subproblems, each projecting the current mean-centered vertex positions $\mathbf{M} \mathbf{V}_{f_i}$ of a face to the corresponding feasible set $\mathcal{C}_i$. Namely, we look for vertex positions $\mathbf{P}_i$ closest to $\mathbf{M} \mathbf{V}_{f_i}$ while achieving the target oriented unit normal $\hat{\mathbf{n}}_i$. The normal condition means that $\mathbf{P}_i$ must lie on a plane orthogonal to $\hat{\mathbf{n}}_i$. Moreover, as the mean of the three vertex positions in $\mathbf{M} \mathbf{V}_{f_i}$ is at the origin, it can be shown that the mean of $\mathbf{P}_i$ must also lie at the origin. As a result, $\mathbf{P}_i$ must lie on a plane that passes through the origin and is orthogonal to $\hat{\mathbf{n}}_i$. The closest projection from $\mathbf{M} \mathbf{V}_{f_i}$ onto this plane can be computed as 
	\[
	\overline{\mathbf{P}}_i = \mathbf{M} \mathbf{V}_{f_i} (\mathbf{I}_3 - \hat{\mathbf{n}}_i \hat{\mathbf{n}}_i^T).
	\] 
	Let $\mathbf{n}_i$ be the oriented unit normal for the current vertex positions $\mathbf{M} \mathbf{V}_{f_i}$. Then depending on the relation between $\mathbf{n}_i$ and $\hat{\mathbf{n}}_i$, we have two possible solutions for $\mathbf{P}_i$. 
	\begin{enumerate}
		\item If $\mathbf{n}_i \cdot \hat{\mathbf{n}}_i \geq 0$, then the oriented unit normal for $\overline{\mathbf{P}}_i$ is $\hat{\mathbf{n}}_i$, and we have $\mathbf{P}_i = \overline{\mathbf{P}}_i$. 
		\item If $\mathbf{n}_i \cdot \hat{\mathbf{n}}_i < 0$, then the oriented unit normal for $\overline{\mathbf{P}}_i$ is $-\hat{\mathbf{n}}_i$. In this case, the solution $\mathbf{P}_i$ degenerates to three colinear points that lie in the plane of $\overline{\mathbf{P}}_i$ and minimizes the distance $\|\mathbf{P}_i - \overline{\mathbf{P}}_i\|_F$. This can be computed as 
		\[
		\mathbf{P}_i = \overline{\mathbf{P}}_i (\mathbf{h} \mathbf{h}^T), 
		\]
		where $\mathbf{h}$ is the right-singular vector of $\overline{\mathbf{P}}_i$ corresponding to its largest singular value.  
	\end{enumerate}
	The subproblem for each face is independent and can be solved in parallel.
	\item Next, we fix $\mathbf{P}$ and update $\mathbf{V}$. This is equivalent to
	\[
	\min_{\mathbf{V}}~~ w \|\mathbf{V} - \mathbf{V}^0\|_F^2 + \|\mathbf{K} \mathbf{V} - \mathbf{P}\|_F^2,
	\]
	where the sparse matrix $\mathbf{K} \in \mathbb{R}^{3 n_f \times n_v}$ collects the mean-centering matrix coefficients for each face. This amounts to solving a sparse positive definite linear system
	\[
	[w \mathbf{I}_{n_v} + \mathbf{K}^T \mathbf{K}] \mathbf{V} = w \mathbf{V}^0 + \mathbf{K}^T \mathbf{P},
	\]
	where $\mathbf{I}_{n_v}$ is the $n_v \times n_v$ identity matrix. The three right-hand-sides of the system corresponds to the $x$-, $y$-, and $z$-coordinates, and can be solved in parallel. Moreover, the system matrix is fixed during all iterations, thus we can pre-compute its Cholesky factorization to allow for efficient solving in each iteration.
\end{itemize}
The above alternating minimization is repeated until convergence. We use 20 iterations in all our experiments, which is sufficient to achieve nice results.

Our vertex update method can efficiently compute a new mesh that is consistent with the target face normals, while being close to the original mesh shape.  Fig.~\ref{fig:VertexUpdateComparison} compares our approach with the vertex update method proposed in~\cite{WangFLTLG15}, which also avoids flipped triangles. 
Given the target oriented normal for each face, they first rotate the current face to align its oriented normal with the target normal; then the new vertex positions are computed by matching the new face gradients with the rotated ones in a least-squares manner, by solving a Poisson linear system.
For each method, we evaluate the deviation between the resulting mesh and the original mesh by aligning their centroids to minimize their $\ell_2$ norm of their vertex deviation (shown in the top center of Fig.~\ref{fig:VertexUpdateComparison}), and then visualizing the deviation of each vertex via color coding. The resulting mesh using our method is noticeably closer to the original mesh, as we explicitly enforce closeness in our target function. This is desirable for many applications such as mesh denoising. In addition, we compute the deviation between the resulting face normals and the target normals, and visualize their distribution using histogram as well as color-coding on the mesh surface. Our method leads to smaller deviation between the target and the resulting normals. Although the computational time of our method (0.4313 seconds) is higher than the method from~\cite{WangFLTLG15} (0.1923 seconds), it does not make a significant difference to the total filtering time (see Table~\ref{table:Time}).

\subsection{SD filter for texture colors}
Our SD filter can be applied to not only face normals, but also other signals defined on mesh surfaces. One example is RGB colors from texture mapping. Given a texture image and its target triangular mesh, we can use the texture coordinates to identify each pixel $i$ that gets mapped to the surface, as well as its target position $\mathbf{p}_i \in \mathbb{R}^3$ on the mesh. In this way, the texture color becomes a signal defined on the mesh surface, associated with the points $\{\mathbf{p}_i\}$. Let $\{\mathbf{g}_i\}$ be the guidance colors for these points. We compute the filtered texture colors $\{\mathbf{q}_i\}$ by minimizing the target function $E_{\textrm{SD}}$ in Eq.~\eqref{eq:objective_energy}, with $\mathbf{c}_i, \mathbf{n}_i$ replaced by $\mathbf{p}_i, \mathbf{q}_i$ respectively, and using the distance between points $\mathbf{p}_i$ to define the neighborhoods and compute the spatial Gaussian weights. Thus the texture colors are filtered according to the metric on the mesh surface instead of the distance on the image plane. 
The optimization problem is solved using unconstrained fixed-point iteration similar to~\eqref{eq:FixedPointIterationFormat}.
Fig.~\ref{fig:TextureRemove-Giraffe} shows some examples of texture color filtering.

\section{Results and Applications}

In this section, we use a series of examples to demonstrate the efficiency and effectiveness of our SD filter, as well as its various applications. We also compare the SD filter with related methods including $\ell_0$ optimization~\cite{he2013mesh} and rolling guidance normal filter (RGNF)~\cite{WangFLTLG15}.

\begin{figure}[t]
	\centering
	\includegraphics[width=1\columnwidth]{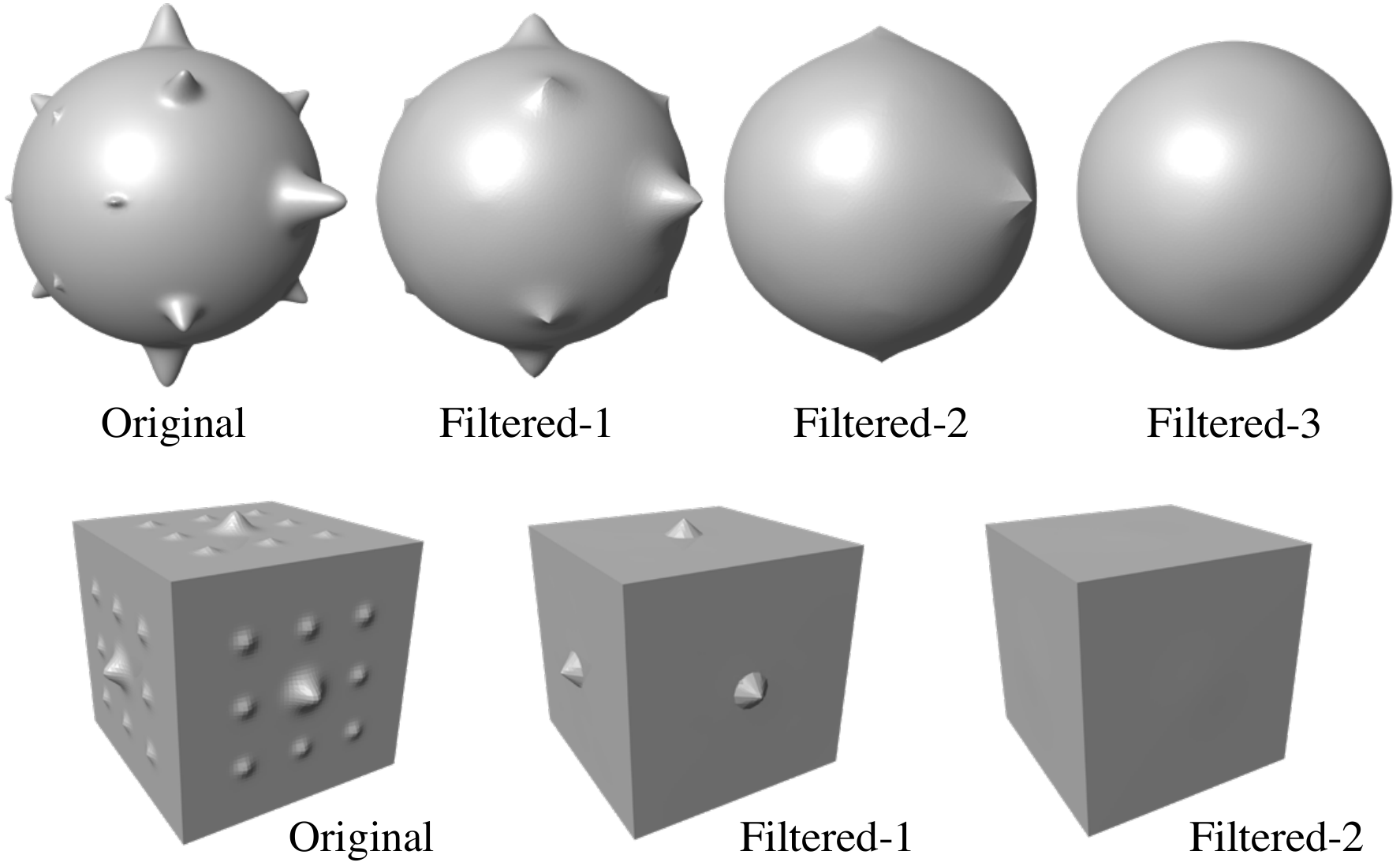}
	\caption{By repeatedly applying our SD normal filter with different parameters, we can gradually remove the geometry features of increasing scales. Detailed parameter values can be found in the supplemental material.}
	\label{fig:TextureRemoval-CubeAndBall}
\end{figure}

\begin{figure}[t]
	\centering
	\includegraphics[width=1\columnwidth]{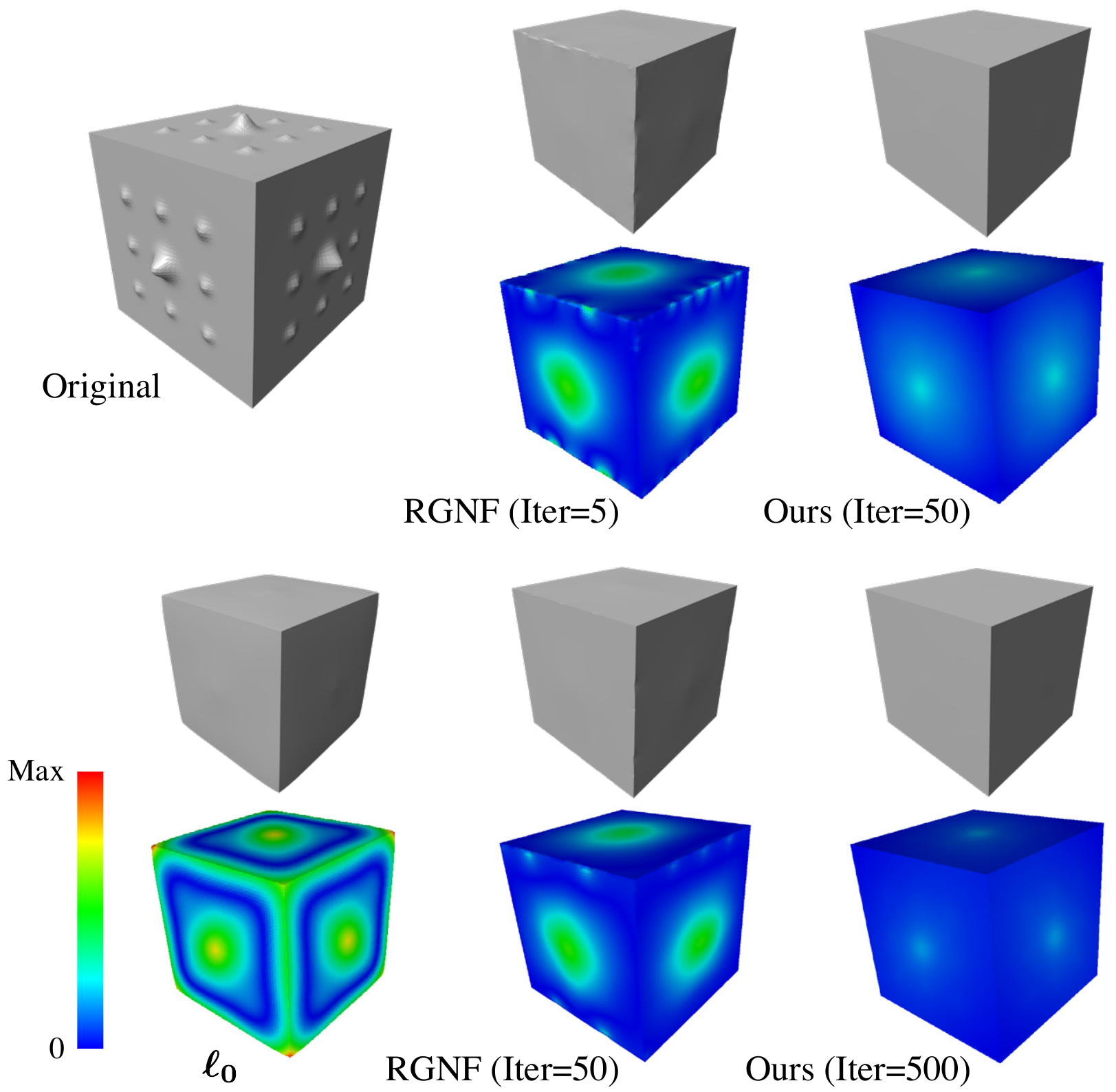}
	\caption{Comparison between the SD normal filter, RGNF, and $\ell_0$ optimization, using a cube shape with added features on each side. The colormap shows the deviation between the results and the original cube shape. The result from the SD filter is the closest to the cube.}
	\label{fig:TextureRemoval-Cube}
\end{figure}

\para{Implementation}
Our algorithm is implemented in C++, using \eigen{}~\cite{eigenweb} for all linear algebra operations. 
For filtering of face normals, we run the iterative solver until one of the following conditions is satisfied: 1) the solver reaches the maximum number iterations, which is set to 100 for all our experiments; or 2) the area-weighted $\ell_2$ norm of normal changes between two consecutive iterations is smaller than a certain threshold angle $\epsilon$, i.e.,
$
\sum_{i} A_i \|\mathbf{n}_i^{k+1} - \mathbf{n}_i^{k}\|^2 \leq 4 \left[\sin(\epsilon / 2)\right]^2\sum_{i} A_i. 
$
We set $\epsilon$ to 0.2 degrees in all our experiments.   
For filtering of texture colors, we run the solver for 50 iterations.
Unless stated otherwise, all examples are run on a desktop PC with 16GB of RAM and a quad-core CPU at 3.6 GHz, using OpenMP for parallelization. 

\begin{figure}[!t]
	\centering
	\includegraphics[width=\columnwidth]{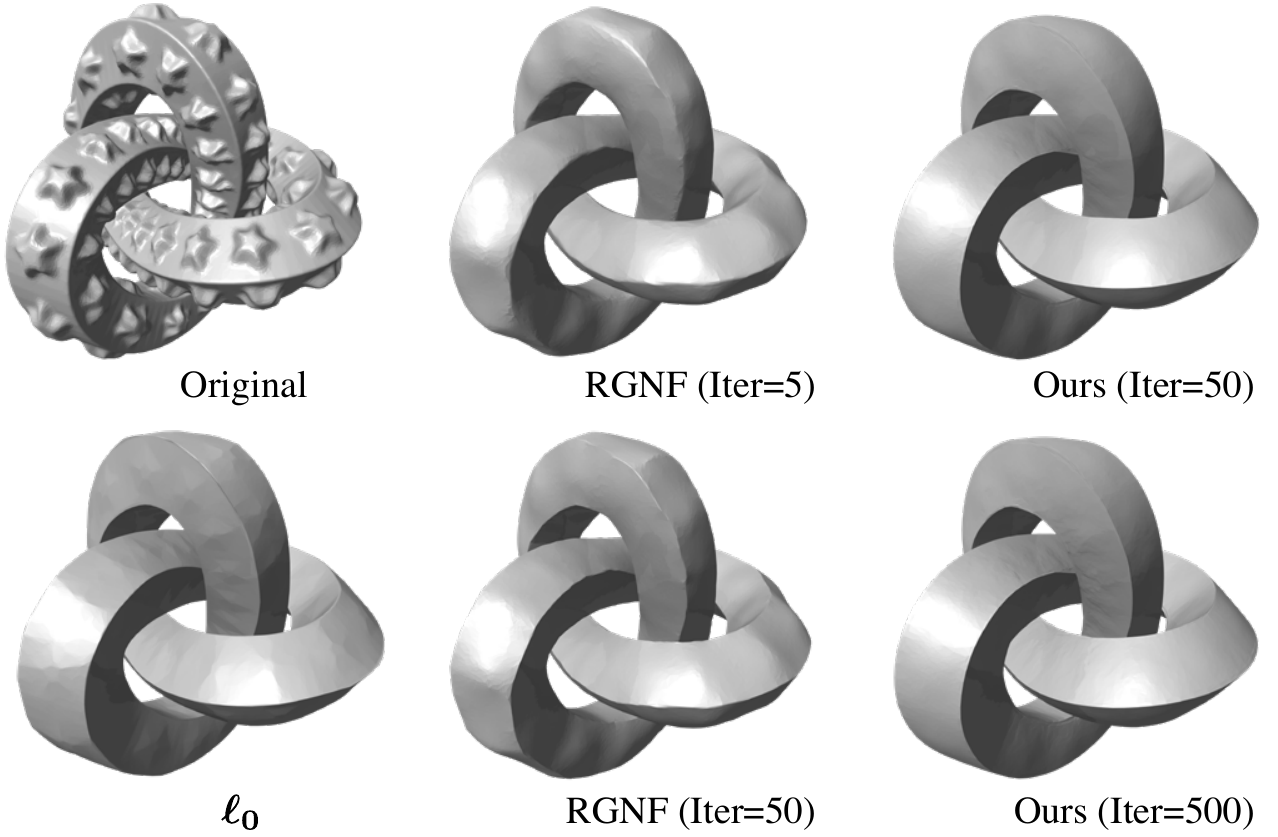}
	\caption{Filtering of the Knot model. The SD filter removes the features on each side of the knot and produces smooth appearance, while sharpening the edges between different sides.}
	\label{fig:TextureRemoval-Knot}
\end{figure}
\begin{figure}[!t]
	\centering
	\includegraphics[width=\columnwidth]{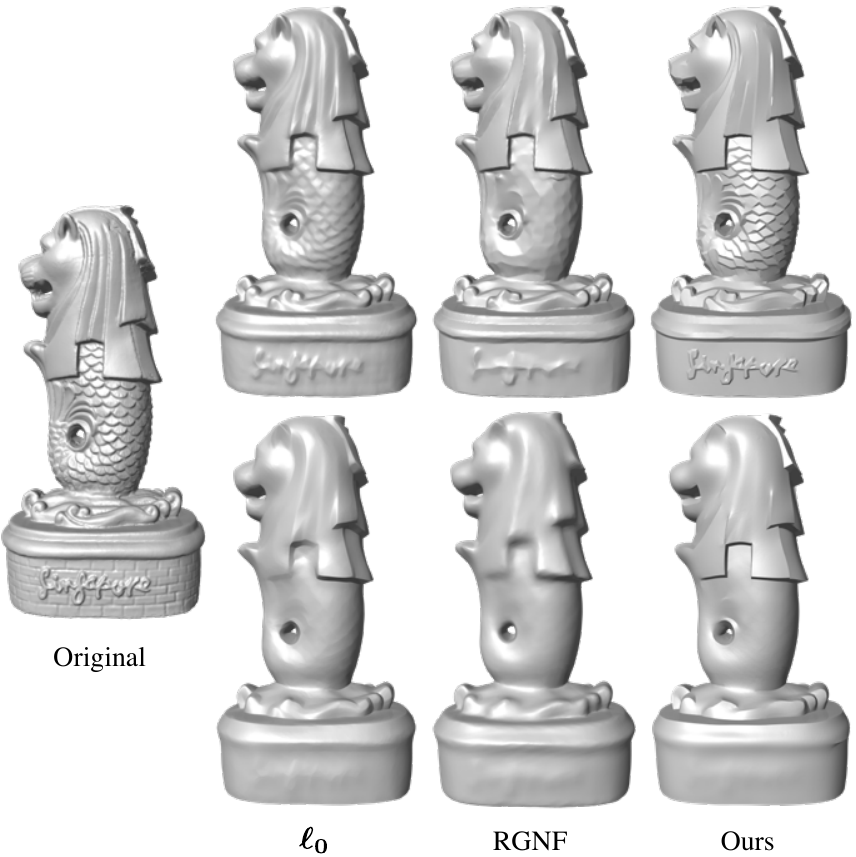}
	\caption{Scale-aware filtering of the Merlion model, in comparison with $\ell_0$ optimization and RGNF. Each row shows comparable results using the three methods, while each column shows the results from one method. The SD filter successfully removes the fine brick lines at the base, while preserving the letters on the base.}
	\label{fig:TextureRemoval-Merlion}
\end{figure}

In all examples, the spatial Gaussian parameter $\eta$ is specified with respect to the average distance between adjacent face centroids, denoted as $l_c$. 
By default, the initial signal is used as the guidance.
For more intuitive control of the optimization, we also rescale the user-specified regularizer weight $\lambda$ according to the value of $\eta$. As $\eta$ increases, the integral of spatial Gaussian $\phi_\eta$ on the corresponding face neighborhood also increases, and the relative scale of the regularizer term with respect to the fidelity term grows. Therefore, we compensate for the change of the regularizer scale due to $\eta$, by rescaling $\lambda$ with a factor $\left(\sum_i A_i\right) / \left(\sum_i \sum_{f_j \in N(i)}  A_{j} \phi_\eta(\|\mathbf{c}_i - \mathbf{c}_j\|) \right)$.

The source code for our SD filter is available at \mbox{\url{https://github.com/bldeng/MeshSDFilter}}. The parameters for the examples in this section can be found in the supplemental material.

\begin{figure}[!t]
	\centering
	\includegraphics[width=0.92\columnwidth]{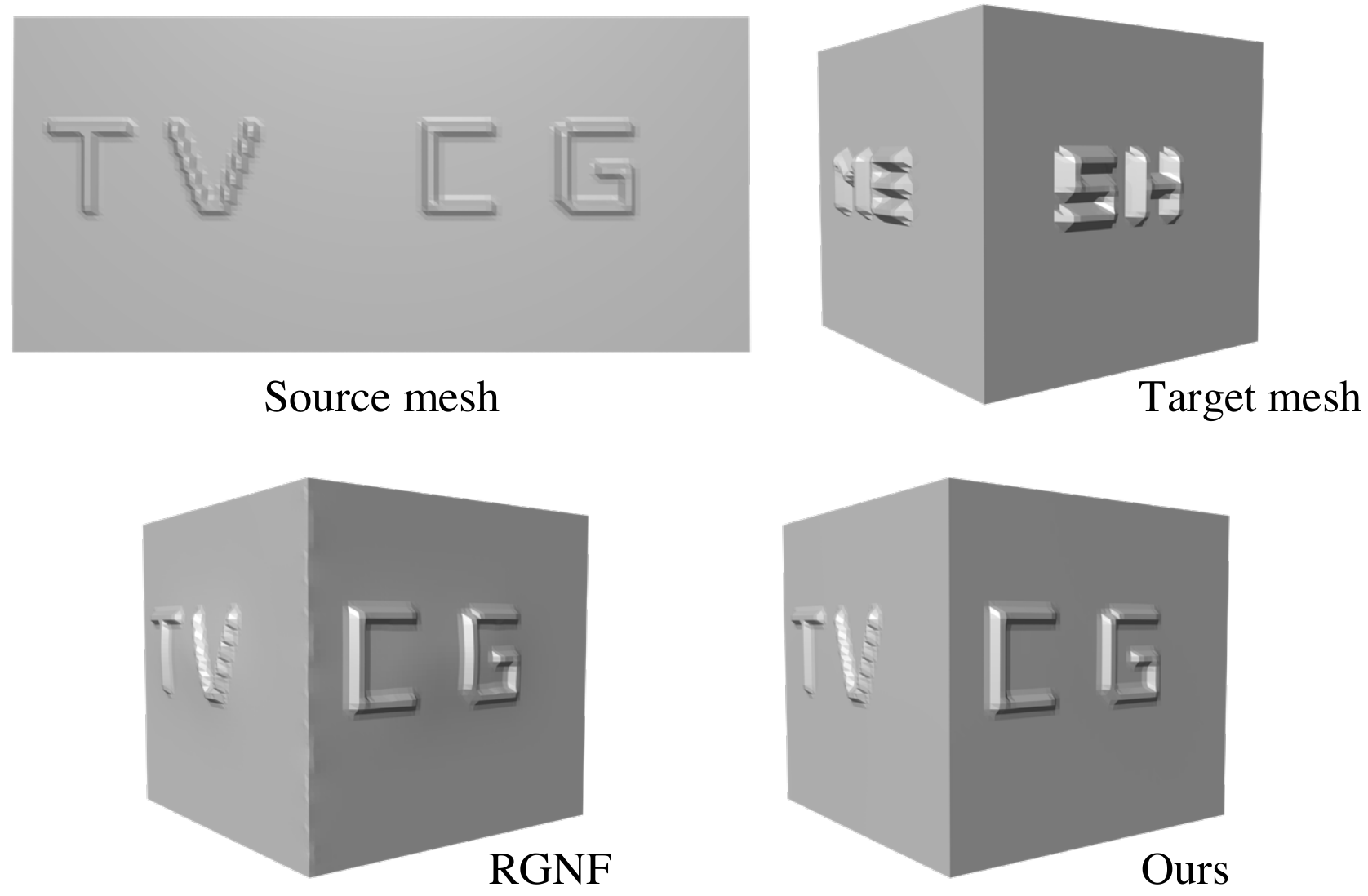}
	\caption{Coating transfer between two meshes, by first filtering the source mesh to obtain its geometric textures, and applying them to the base cube shape of the target mesh which is also obtained via filtering. The SD filter produces a better result in preserving the sharp edges and recovering the flat faces of the underlying cube shape.}
	\label{fig:CoatingTransfer}
\end{figure}

\para{Scale-aware and feature-preserving filtering}
The SD filter can effectively remove features based on their scales according to the user-specified parameters. This is demonstrated in Fig.~\ref{fig:TextureRemoval-CubeAndBall}, where the input models are a sphere and a cube with additional features of different scales on the surfaces. Using different parameter settings, the SD filter gradually removes the geometry features of increasing scales, while preserving the sharp edges on the cube. Similar scale-aware and feature-preserving effects are observed for filtering of texture colors, as shown in Fig.~\ref{fig:TextureRemove-Giraffe}.

In Figs.~\ref{fig:TextureRemoval-Cube}, \ref{fig:TextureRemoval-Knot}, and \ref{fig:TextureRemoval-Merlion}, we show more examples of scale-aware and feature-preserving filtering of mesh geometry using the SD filter, and compare the results with $\ell_0$ optimization method from~\cite{he2013mesh} and RGNF from~\cite{WangFLTLG15}.  
We tune the parameters of each method to achieve the best results, while ensuring comparable effects from different methods.  
In all examples, the SD filter achieves better or similar results compared with RGNF, and outperforms $\ell_0$ optimization. In Fig.~\ref{fig:TextureRemoval-Cube}, the input model is a cube with additional features on each face, and the result with the SD filter is the closest to the cube shape. The result with RGNF has larger deviation from the cube shape, because the filtered signals are computed as a combination of the original signals within a neighborhood; as a consequence, when there is large deviation between the input signals and the desired output within a certain region, RGNF may not produce a desirable result inside the region. 
By default, Wang et al.~\cite{WangFLTLG15} run RGNF for 5 iterations. To verify its convergence, we run 50 iterations of RGNF and compare the results. We can see that the resulting mesh with 50 RGNF iterations is slightly closer to the cube shape, but the difference is not significant, and the undesirable deviation around the edges remains. For comparison, we also run the SD filter for 50 and 500 iterations respectively. Similar to RGNF, running more iterations of SD filters makes the result slightly closer to the cube shape, but without significant difference. 
In Fig.~\ref{fig:TextureRemoval-Knot}, the SD filter is able to smooth out the star-shaped features on the knot surface, while enhancing the sharp feature lines between different sides of the knot. Although $\ell_0$ optimization also enhances the feature lines, it leads to piecewise flat shapes because the $\ell_0$ norm promotes piecewise constant signals.
Similar to Fig.~\ref{fig:TextureRemoval-Cube}, we run RGNF for 5 and 50 iterations, and the SD filter for 50 and 500 filters, to compare their results and convergence. For both filters, running for more iterations slightly improve the result, but without significant difference.
In Fig.~\ref{fig:TextureRemoval-Merlion}, the three methods produce similar results on the Merlion model, while the scale-awareness of the SD filter enables it to remove the fine brick lines at the base while clearly retaining the letters.

\begin{figure}[t]
	\centering
	\includegraphics[width=1\columnwidth]{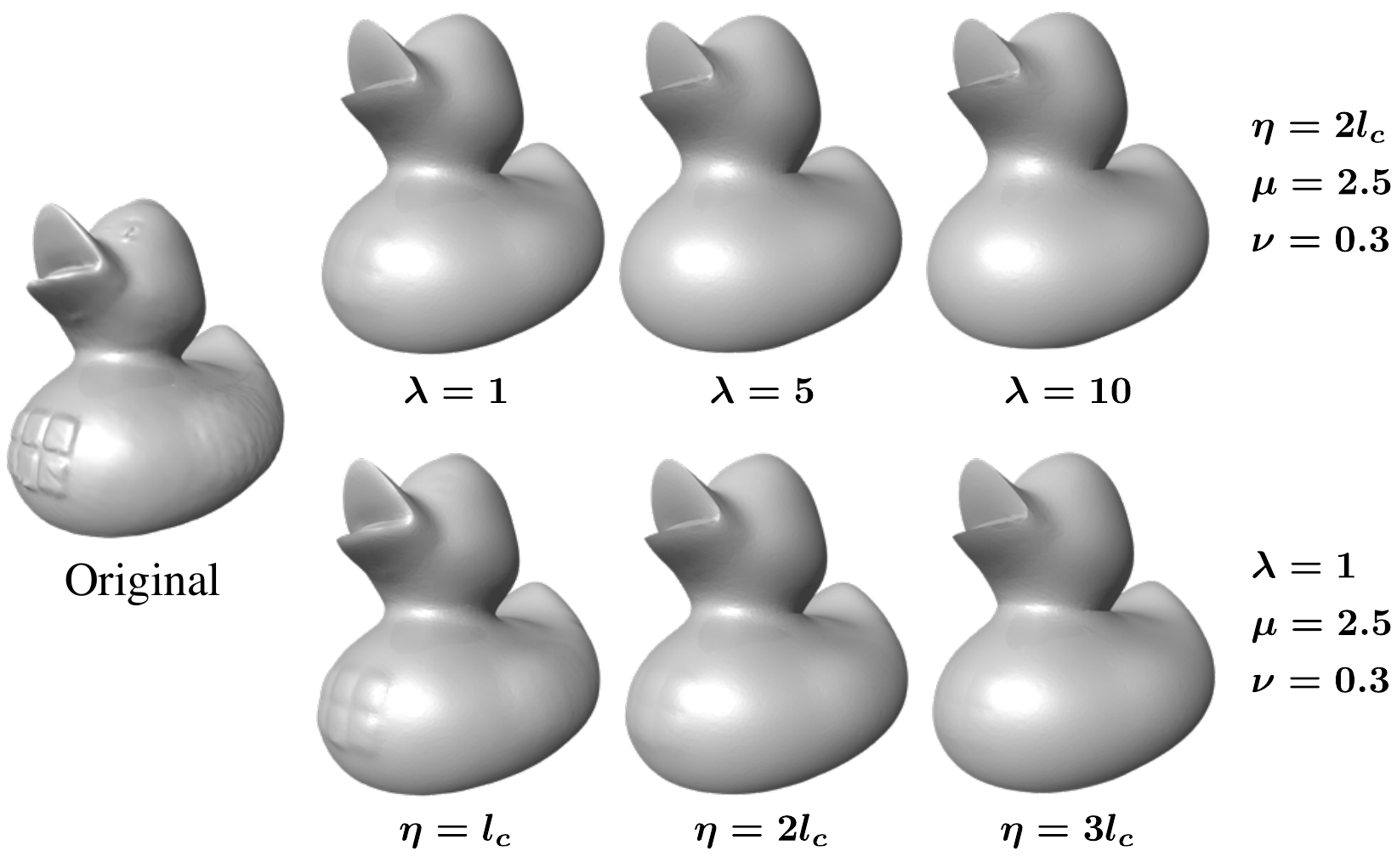}
	\caption{A larger value of $\lambda$ or $\eta$ leads to a smoother filtering result. Here each rows shows the filtering results with increasing values of $\lambda$ or $\eta$, while keeping the other parameters fixed.}
	\label{fig:parameters}
\end{figure}

\begin{figure*}[t]
	\centering
	\includegraphics[width=\textwidth]{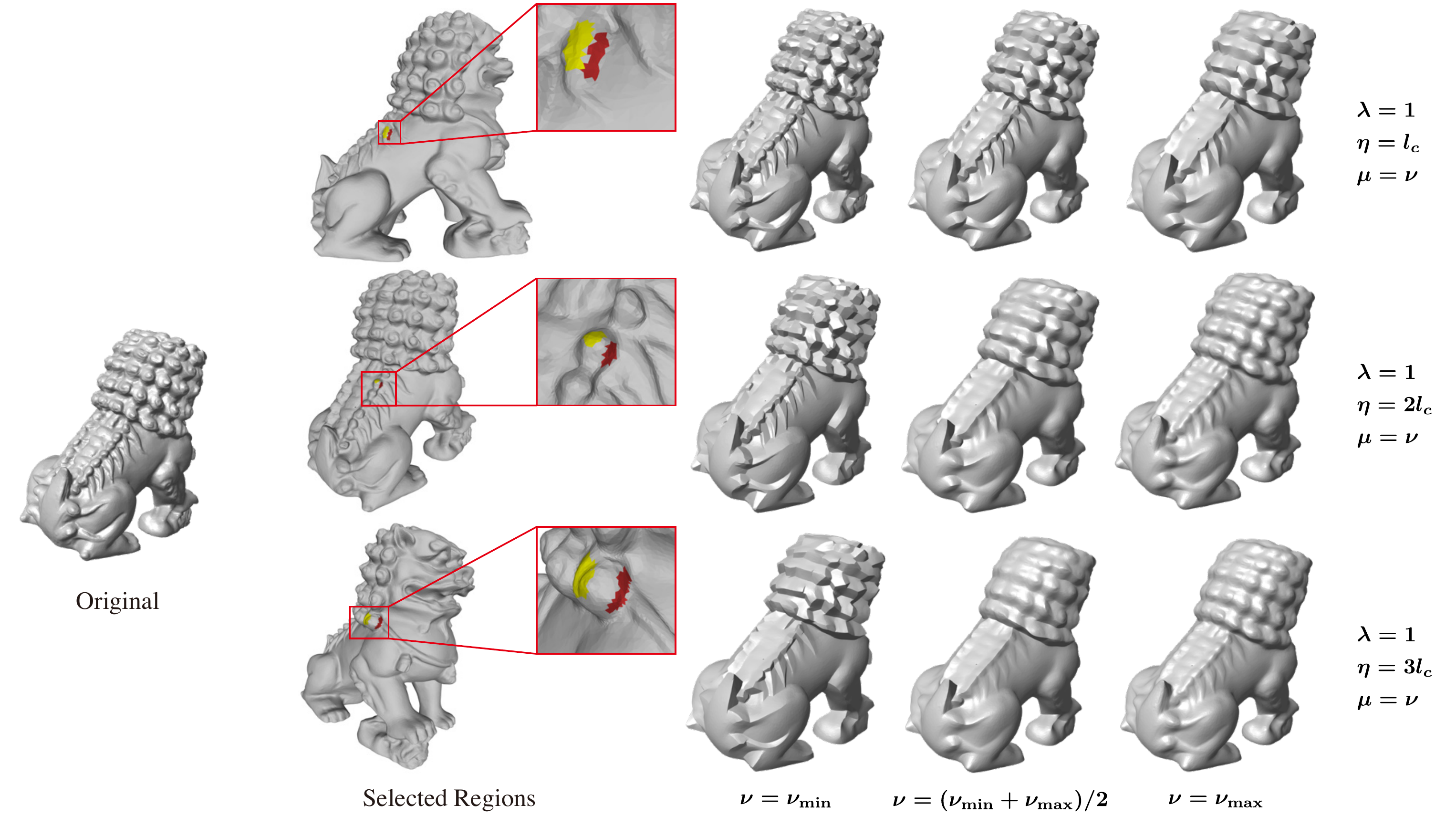}
	\caption{According to two user-selected regions (shown in red and yellow), we determine upper bound $\nu_{\max{}}$ and $\nu_{\min{}}$ for the $\nu$ parameter, which inhibits influence between the face normals from the two regions, and ensures the smoothing of normals within each region. Within this range, a larger $\nu$ leads to smoother results, while a smaller $\nu$ promotes sharp features between the two selected regions. Here each row shows one pair of selected regions, and the resulting meshes using different values of $\nu$ within the corresponding range.}
	\label{fig:ParameterSelectionBasedUserInteraction-Lion}
\end{figure*}

\para{Coating transfer}
Similar to~\cite{WangFLTLG15}, our filter can be utilized to perform coating transfer~\cite{Sorkine2004}, which transfers geometric textures from a source mesh to a target mesh. An example is shown in Fig.~\ref{fig:CoatingTransfer}. Here following~\cite{WangFLTLG15}, we first filter the source mesh to remove the letters on its surface, and encode them using the difference between Laplacian coordinates of the mesh before and after the filtering. We then transfer these letters onto the base cube shape of the target mesh which is also obtained by filtering: the encoded letter shapes are added to the Laplacian coordinates of the base cube mesh, and a new mesh is reconstructed from the resulting Laplacian coordinates. For comparison, we also show the results using~\cite{WangFLTLG15}, where the meshes are processed using the RGNF instead of the SD filter. While both methods can extract the source geometric textures and prepare the target cube mesh, the SD filter produces better a result in preserving the sharp edges and recovering the flat faces of the cube shape.

\para{Choice of parameters}
Our filter is influenced by four parameters: the regularizer weight $\lambda$, the spatial Gaussian parameter $\eta$ which also influences the neighborhood size, the guidance Gaussian parameter $\mu$, and the range Gaussian parameter $\nu$. For both $\lambda$ and $\eta$, a larger value leads to a smoother result, as shown in Fig.~\ref{fig:parameters}. Parameters $\mu$ and $\nu$ determine which face normals within a neighborhood affects the central face in a fixed-point iteration. 
For more intuitive control, we propose a method to interactively set the $\mu$ and $\nu$ parameters. Firstly, the user selects two smooth regions on different sides of a feature intended to be kept. We denote the two regions as $P_1$ and $P_2$, and compute the mean $\widetilde{\mathbf{n}}_l$ and variance $\sigma_l$ ($l=1,2$) of the normals within each region via
\[
{\widetilde{\mathbf{n}}_{l}} = \frac{\sum_{i\in {{P}_{l}}}{{{A}_{i}}{\mathbf{n}_{i}}}}{\left\| \sum_{i\in {{P}_{l}}}{{{A}_{i}}{\mathbf{n}_{i}}} \right\|},  
\quad
\sigma_l^2
= \frac{\sum_{i \in P_l} A_i \left\| \mathbf{n}_i - \widetilde{\mathbf{n}}_l \right\|^2}
{\sum_{i \in P_l} A_i}.
\]
Then the range of $\nu$ is determined using the following strategy: $\nu$ should be small enough such that the two mean normals have negligible influence on each other according to the range Gaussian; at the same time, within each region the normals should influence each other such that sharp features do not emerge. Based on this strategy, we first determine the lower- and upper-bounds of $\nu$ via
\[
\nu_{\min} = \frac{1}{2}\max(\sigma_1, \sigma_2),
\quad
\nu_{\max} = \frac{1}{3}\left\|\widetilde{\mathbf{n}}_1 - \widetilde{\mathbf{n}}_2\right\|.
\]
If $\nu_{\max} < \nu_{\min}$, then the user needs to select another pair of regions; otherwise, a value between $\nu_{\min}$ and $\nu_{\max}$ as the parameter $\nu$. In our experiments, good results can often be achieved by choosing $\nu = (\nu_{\min} + \nu_{\max}) / 2$ and setting $\mu = a \cdot \nu$ with $a \in [1, 10]$.
Fig~\ref{fig:ParameterSelectionBasedUserInteraction-Lion} shows the effects of different $\nu$ values on the Chinese lion model.

 \begin{figure*}[!t]
 	\centering
 	\includegraphics[width=\textwidth]{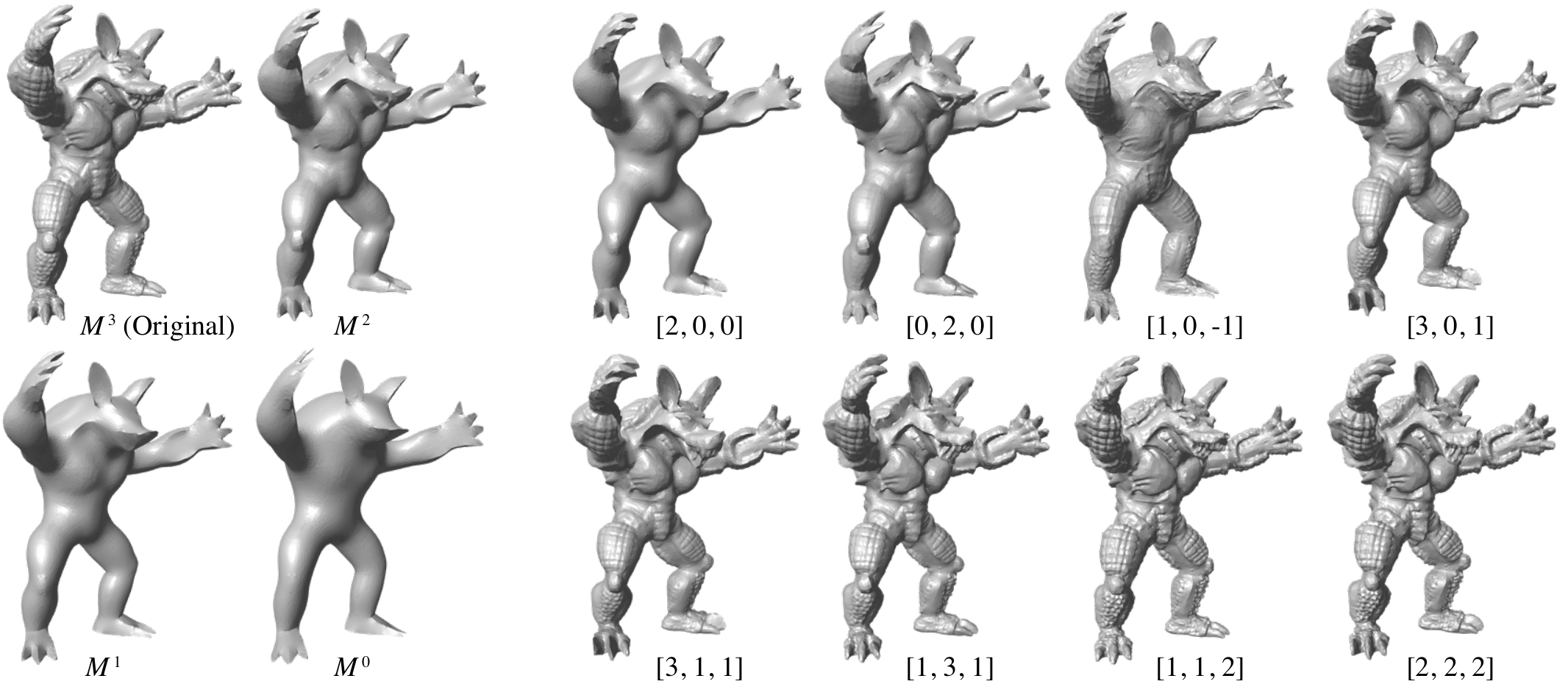}
 	\caption{Geometry feature manipulation and enhancement for the Armadillo model, by controlling the contribution from features of different scales. Left: a coarse-to-fine sequence of meshes, obtained by repeatedly applying the SD normal filter with different parameters. Right: new meshes generated using linearly combined target vertex positions and target normals (Eq.~\eqref{eq:FeatureCombination}), with the combination coefficients shown below each result.}
 	\label{fig:Armadillo}
 \end{figure*}
 
 \begin{figure*}[!t]
 	\centering
 	\includegraphics[width=\textwidth]{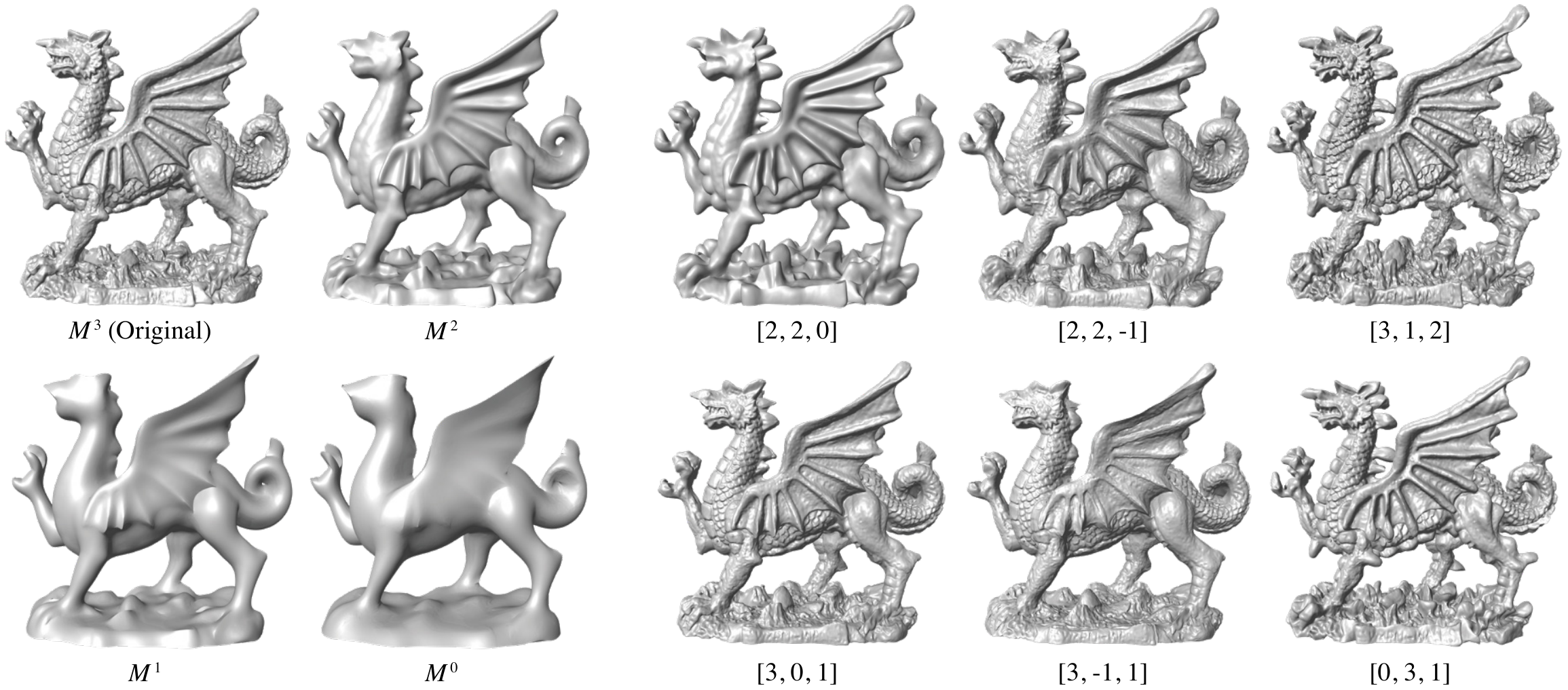}
 	\caption{Geometry feature manipulation and enhancement for the Welsh Dragon model. Left: the coarse-to-fine sequence of meshes resulting from SD normal filtering. Right: generated new meshes and their corresponding linear combination coefficients.}
 	\label{fig:WelshDragon}
 \end{figure*}

\begin{figure*}[!t]
	\centering
	\includegraphics[width=\textwidth]{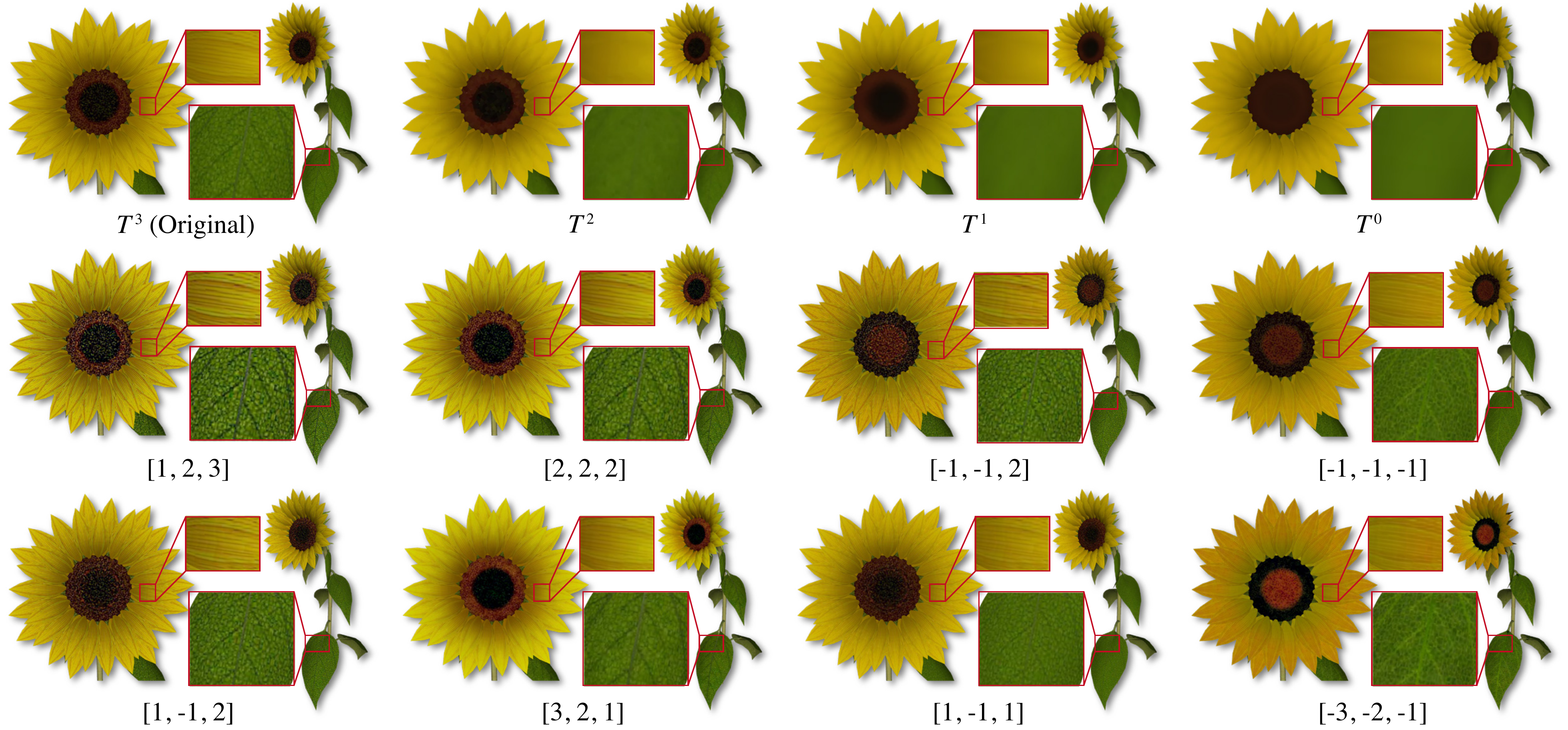}
	\caption{Texture image feature enhancement using the SD filter.}
	\label{fig:ImageTextureEditting-Sunflower}
\end{figure*}

\begin{figure}[!b]
	\centering
	\includegraphics[width=1\columnwidth]{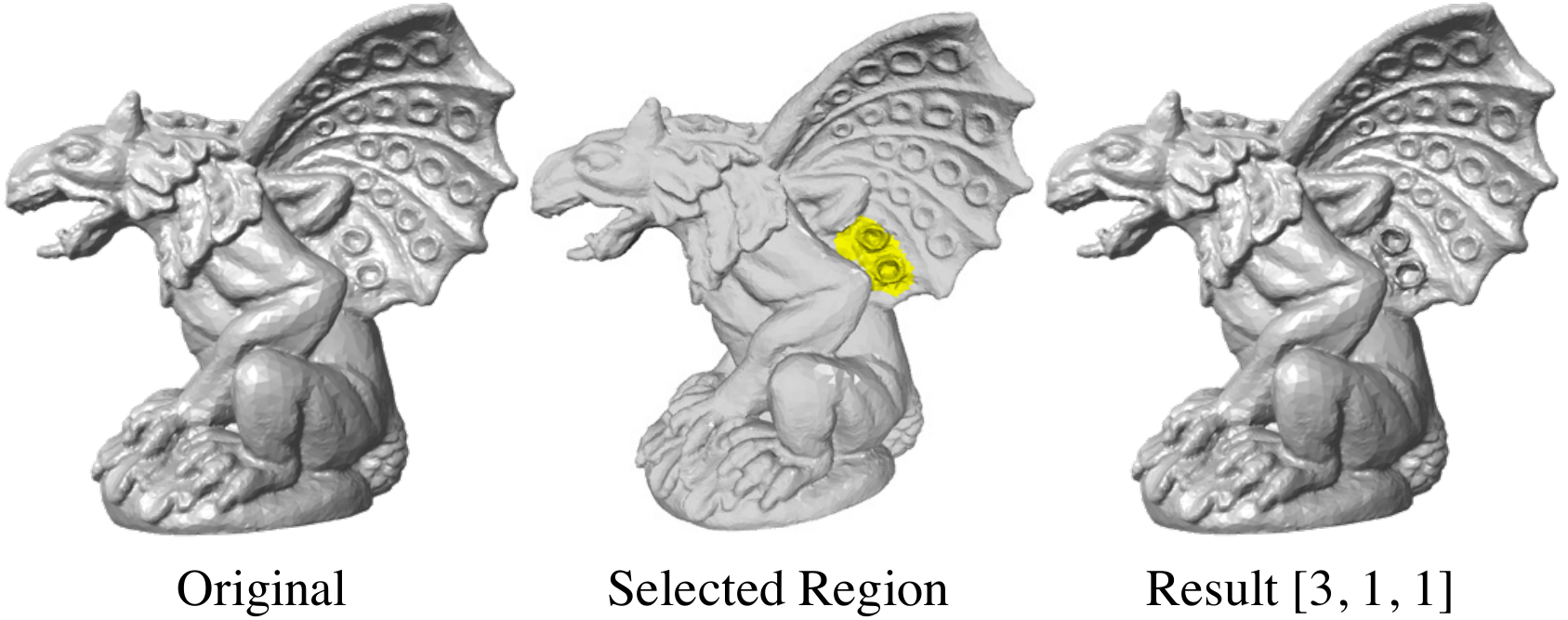}
	\caption{Detail enhancement of the Gargoyle model in a local region selected by the user (shown in yellow), using the same coarse-to-fine mesh sequence as Fig.~\ref{fig:teaser}.}
	\label{fig:GeometryTextureEditting-Selective}
\end{figure}

\begin{figure}[!b]
	\centering
	\includegraphics[width=\columnwidth]{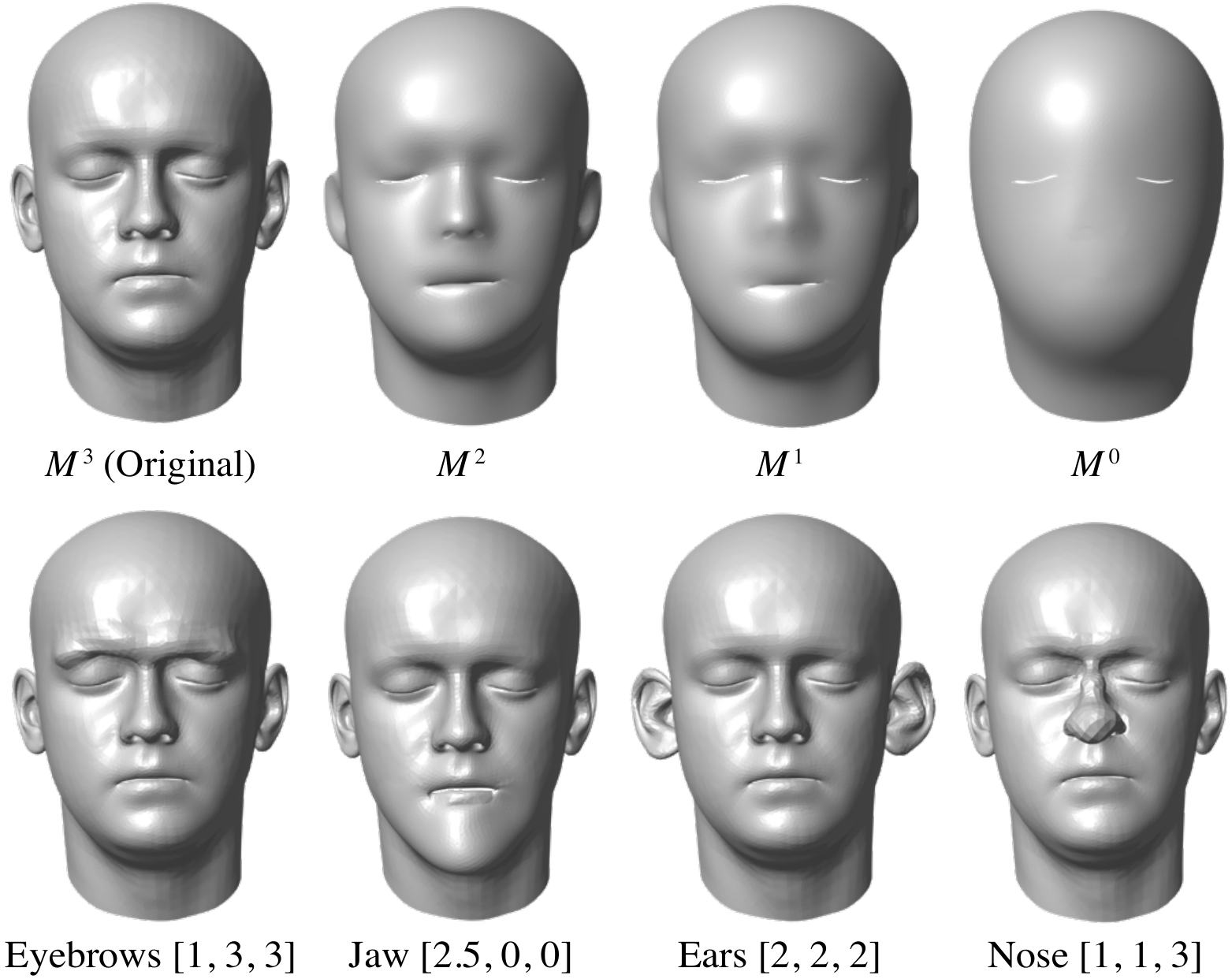}
	\caption{Local feature enhancement on a human face model, with the local region and the combination coefficients annotated below each result.}
	\label{fig:GeometryTextureEditting-FaceLocal}
\end{figure}

\para{Feature manipulation and enhancement}
The scale-awareness of our filter enables us to manipulate mesh details according to their scales. Given an input mesh $M$, we repeatedly apply the SD normal filter with different parameters to obtain a series of filtered meshes $M^{m-1}, \ldots, M^{1}, M^{0}$, where $M^k (k=0,\ldots,m-2)$ is obtained by filtering $M^{k+1}$ to remove more details. If we denote the input mesh as $M^m$, then $M^0, \ldots, M^m$ forms a coarse-to-fine sequence of meshes, with $M^0$ being the \emph{base mesh}, and $M^m$ being the original mesh. We encode the difference between two consecutive meshes $M^{k}, M^{k-1}$ by comparing their corresponding vertex positions and face normals, represented as $\mathbf{d}_i^k = \mathbf{v}_i^{(k)} - \mathbf{v}_i^{(k-1)} (i = 1, \ldots, n_v)$ and $\mathbf{f}_j^k = \mathbf{n}_j^{(k)} - \mathbf{n}_j^{(k-1)} (i = 1, \ldots, n_f)$, where $\left\{\mathbf{v}_i^{(k)}\right\}$ and $\left\{\mathbf{n}_j^{(k)}\right\}$ are the vertex positions and face normals of mesh $M^k$. These differences represent the required deformation for $M^{k-1}$ to introduce the additional details in $M^k$. They can be linearly combined according to coefficients $\boldsymbol{\alpha} = [\alpha_1, \ldots, \alpha_m]$ and added to the face normals and vertex positions of the base mesh, to derive the target vertex positions $\left\{\hat{\mathbf{v}}_i\right\}$ and target face normals $\left\{\hat{\mathbf{n}}_j\right\}$ for a new mesh $\overline{M}$:
\begin{equation}
\hat{\mathbf{v}}_i = \mathbf{v}_i^{(0)} + \sum_{k=1}^m \alpha_k \mathbf{d}_i^k, \quad 
\hat{\mathbf{n}}_j = \frac{\mathbf{n}_j^{(0)} + \sum_{k=1}^m \alpha_k \mathbf{f}_j^k}
{\|\mathbf{n}_j^{(0)} + \sum_{k=1}^m \alpha_k \mathbf{f}_j^k\|}.
\label{eq:FeatureCombination}
\end{equation}
Note that the target vertex positions and target face normals are often incompatible. To combine the two conditions, we determine the new mesh by solving the same optimization problem as our vertex update~\eqref{eq:VertexUpdateTargetFunction}, with the matrix $\mathbf{V}^0$ in the target function storing the target vertex positions. In this way, the linear combination coefficients $\boldsymbol{\alpha}$ indicate the contribution of geometric features from the original model within a certain range of scales. By changing the value of $\boldsymbol{\alpha}$, a user can control geometric features according to their scales. Setting all coefficients to 1 recovers the original mesh, while setting a coefficient to a value different from 1 can boost or attenuate the features of the corresponding scales. Moreover, the linear system matrix for the optimization problem is fixed regardless of the value of $\boldsymbol{\alpha}$, and only needs to be pre-factorized once. Afterwards, the user can choose any value of $\boldsymbol{\alpha}$, and the resulting mesh can be efficiently computed using the pre-factorized system. This allows the user to interactively explore different linear combination coefficients to achieve desirable results. Figs.~\ref{fig:teaser}, \ref{fig:Armadillo}, and \ref{fig:WelshDragon} show examples of new meshes created in this manner. We can see that the coarse-to-fine sequence of meshes captures the geometrical features of different scales, which are effectively manipulated using the linear combination coefficients. In some application scenarios, it is desirable to only modify the features within a certain region on the surface. In this case, the target vertex positions and target normals are linearly combined only within user-selected regions; outside these regions they remain the same as the original mesh. Fig.~\ref{fig:GeometryTextureEditting-Selective} shows such an example, where we enhance the features of the Gargoyle model in a user-selected local region. Another example is shown in Fig.~\ref{fig:GeometryTextureEditting-FaceLocal}, where a 3D human face model is locally enhanced.

Similarly, we can manipulate and enhance texture colors using the SD filter, as shown in Fig.~\ref{fig:ImageTextureEditting-Sunflower}. We first filter the input texture colors $T$ incrementally to derive a coarse-to-fine sequence $T^{0}, \ldots, T^{m-1}, T^{m} = T$. Then new texture colors are computed via linear combination with coefficients $\boldsymbol{\alpha}$:
\begin{equation}
{T}_{\textrm{new}}={T^{0}}+\sum_{k=1}^{m}{{\alpha }_{k}}(T^k - T^{k-1}).
\end{equation}
Out-of-bound colors in ${T}_{\textrm{new}}$ are reset to their closest valid values.

\begin{figure}[!b]
	\centering
	\includegraphics[width=1\columnwidth]{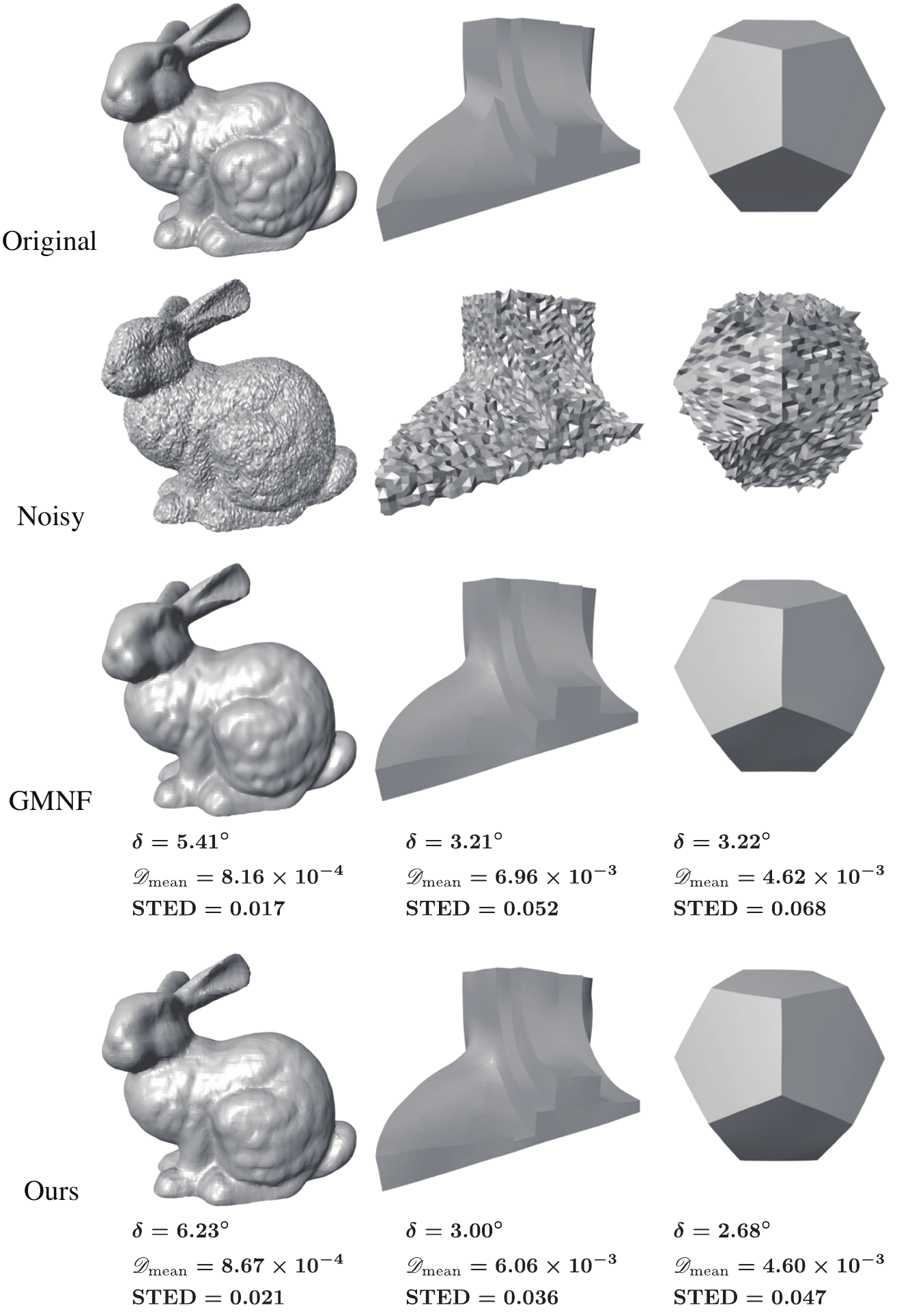}
	\caption{Comparison between denoising results using the SD normal filter, and the guided mesh normal filtering (GMNF) method~\protect\cite{ZhangDZBL15}. The results from the two methods are close, with similar error metric values.}
	\label{fig:Denoising}
\end{figure}

\para{Mesh denoising}
By constructing appropriate guidance signals, our SD filter can also be applied for mesh denoising, as shown in Fig.~\ref{fig:Denoising}. Here we repeatedly apply the SD normal filter to each input mesh to remove the noise. In each run of the filter, the guidance normals are computed from the current mesh using the patch-based approach proposed in~\cite{ZhangDZBL15}. The results are evaluated using the average normal deviation $\meanangleerror$ and the average vertex deviation $\meandist$ from the ground-truth mesh as proposed in~\cite{ZhangDZBL15}. In addition, similar to~\cite{solomon2014general}, we measure the perceptual difference between the denoised mesh and the ground truth using the spatial error term of the STED distance proposed in~\cite{Vasa2011}, computed with one-ring vertex neighborhood. Fig.~\ref{fig:Denoising} compares our denoising results with the guided mesh normal filtering (GMNF) method from~\cite{ZhangDZBL15}. The results from the two methods are quite close, with similar error metric values. Detailed parameter settings are provided in the supplementary materials.

\begin{table*}[t]
	\caption{Computational Time (in Seconds) for Different Parts of the SD Normal Filtering Method}
	\label{table:Time}
	\centering
	\begin{tabular}{ | c | c | c | c | c | c | c | c | c | c |}
		\hline
		Model & \#Faces  & $\lambda$ & $\eta$ & $\mu$
		& $\nu$ & $T_1$ & $T_2$ & $T_3$ & $T_{\textrm{total}}$ \\
		\hline
		
		Armadillo &
		43K& 2 & $2.5 l_c$  & 1.5 & 0.45 &  0.57 &0.036& 0.19 & 1.45   \\ \cline{1-10}

		Cube&
		49K & $ 1 \times 10^6$ & $5l_c$& 2.5 &  0.4& 2.03 & 0.15 & 0.37 & 3.42 \\ \cline{1-10}
		Sphere &
		60K& 100 & $3 l_c$& 1.5 &  0.17& 1.10 & 0.065 & 0.34 & 2.88   \\ \cline{1-10}
		Duck&
		68K& 10 & $2 l_c$& 2.5 &  0.3 & 0.61 & 0.031 & 0.33 & 1.69   \\ \cline{1-10}

		Knot &
		100K& $10$ & 2$l_c$&  2.5 & 0.8& 1.30 & 0.07 & 0.50& 5.69\\ \cline{1-10}
		
		Gargoyle&
		100K& 5 & $3 l_c$&  10 & 0.42 & 1.90 & 0.12 & 0.35 & 8.17  \\ \cline{1-10}
		Merlion &
		566K & 10 & $2.5 l_c$&  2 & 0.26& 6.96 & 0.49 & 5.24 & 32.22   \\ \cline{1-10}
		Welsh Dragon &
		2.21M & 10 & $1.5 l_c$&  20 & 0.35& 19.85 & 0.49 & 36.17 & 71.26   \\ \cline{1-10}
		\hline
	\end{tabular}
\end{table*}

\para{Performance}
Using parallelization, our SD filter can compute the results efficiently. Table~\ref{table:Time} provides the representative computation time of the SD normal filter on different models, showing the timing for each part of the algorithm:
\begin{itemize}[leftmargin=*]
	\item $T_1$: the pre-processing time for finding the neighborhood; 
	\item $T_2$: the average timing of one iteration;
	\item $T_3$: the timing for mesh vertex update;
	\item $T_{\textrm{total}}$: The total timing for the whole filtering process.
\end{itemize}
For meshes with less than 100K faces and with $\eta \leq 3 l_c$, the whole process typically takes only a few seconds.

\section{Discussion and Conclusion}

We present the SD filter for triangular meshes, which is formulated as an optimization problem with a target energy that combines a quadratic fidelity term and a nonconvex robust regularizer. We develop an efficient fixed-point iteration solver for the problem, enabling the filter to be applied for interactive applications. Our SD filter generalizes the joint bilateral filter, combining the static guidance with a dynamic guidance that is derived from the current signal values. Thanks to the joint static/dynamic guidance, the SD filter is robust, feature-preserving and scale-aware, producing state-of-the-art results for various geometry processing problems.

Although our solver can incorporate simple constraints such as unit length for normal vectors, we do not consider global conditions for the signals.
For example, we do not ensure the integrability of normals, i.e., the existence of a mesh whose face normals match the filter results; as a result, some parts of the updated mesh may not be consistent with the filtered normals. Neither do we consider the prevention of self intersection of the updated mesh. Due to the local nature of our fixed-point iteration, it is not easy to incorporate such global constraints into the solver. A possible remedy is to introduce a separate step to enforce these conditions after a few iterations. A more in-depth investigation into such global conditions will be an interesting problem.

Depending on the parameters, our filter can sometimes create sharp features in regions that are originally smooth. For example, this can be seen in the top right image of Fig.~\ref{fig:TextureRemoval-Merlion}, around the scales of the Merion model. One potential solution is to use spatially varying filter parameters that adapt to local feature sizes and user preferences, which we will leave as future work.

Although we only consider filtering of face normals and texture colors in this paper, our formulation is general enough to be applied for other scenarios. In the future, we would like to extend the filter to other geometry signals such as curvatures and shape operators, and to other geometric representations such as point clouds and implicit surfaces. Also worth investigating is the application of fixed-point iteration to similar image filtering problems, especially where direct linear solve is too slow due to the problem scale or the neighborhood size.

\section*{Acknowledgments}
We thank Yang Liu for providing the implementation of RGNF. The Welsh Dragon mesh model was released by Bangor University, UK, for Eurographics 2011. This work was supported by the National Key R\&D Program of China (No. 2016YFC0800501), the National Natural Science Foundation of China (No. 61672481, No. 61672482 and No. 11626253), and the One Hundred Talent Project of the Chinese Academy of Sciences.

\bibliographystyle{IEEEtran}
\bibliography{MeshSDFilter}

\begin{thebibliography}{10}
\providecommand{\url}[1]{#1}
\csname url@samestyle\endcsname
\providecommand{\newblock}{\relax}
\providecommand{\bibinfo}[2]{#2}
\providecommand{\BIBentrySTDinterwordspacing}{\spaceskip=0pt\relax}
\providecommand{\BIBentryALTinterwordstretchfactor}{4}
\providecommand{\BIBentryALTinterwordspacing}{\spaceskip=\fontdimen2\font plus
\BIBentryALTinterwordstretchfactor\fontdimen3\font minus
  \fontdimen4\font\relax}
\providecommand{\BIBforeignlanguage}[2]{{%
\expandafter\ifx\csname l@#1\endcsname\relax
\typeout{** WARNING: IEEEtran.bst: No hyphenation pattern has been}%
\typeout{** loaded for the language `#1'. Using the pattern for}%
\typeout{** the default language instead.}%
\else
\language=\csname l@#1\endcsname
\fi
#2}}
\providecommand{\BIBdecl}{\relax}
\BIBdecl

\bibitem{bilateral1998}
C.~Tomasi and R.~Manduchi, ``Bilateral filtering for gray and color images,''
  ser. ICCV '98, 1998.

\bibitem{eisemann2004flash}
E.~Eisemann and F.~Durand, ``Flash photography enhancement via intrinsic
  relighting,'' \emph{ACM Trans. Graph.}, vol.~23, no.~3, pp. 673--678, 2004.

\bibitem{petschnigg2004digital}
G.~Petschnigg, R.~Szeliski, M.~Agrawala, M.~Cohen, H.~Hoppe, and K.~Toyama,
  ``Digital photography with flash and no-flash image pairs,'' \emph{ACM Trans.
  Graph.}, vol.~23, no.~3, pp. 664--672, 2004.

\bibitem{cho2014bilateral}
H.~Cho, H.~Lee, H.~Kang, and S.~Lee, ``Bilateral texture filtering,'' \emph{ACM
  Trans. Graph.}, vol.~33, no.~4, pp. 128:1--128:8, 2014.

\bibitem{ZhangSXJ14}
Q.~Zhang, X.~Shen, L.~Xu, and J.~Jia, ``Rolling guidance filter,'' in
  \emph{Computer Vision--ECCV 2014}.\hskip 1em plus 0.5em minus 0.4em\relax
  Springer, 2014, pp. 815--830.

\bibitem{fleishman2003bilateral}
S.~Fleishman, I.~Drori, and D.~Cohen-Or, ``Bilateral mesh denoising,''
  \emph{ACM Trans. Graph.}, vol.~22, no.~3, 2003.

\bibitem{jones2003non}
T.~R. Jones, F.~Durand, and M.~Desbrun, ``Non-iterative, feature-preserving
  mesh smoothing,'' \emph{ACM Trans. Graph.}, vol.~22, no.~3, pp. 943--949,
  2003.

\bibitem{zheng2011bilateral}
Y.~Zheng, H.~Fu, O.-C. Au, and C.-L. Tai, ``Bilateral normal filtering for mesh
  denoising,'' \emph{IEEE Trans. Vis. Comput. Graphics}, vol.~17, no.~10, pp.
  1521--1530, 2011.

\bibitem{solomon2014general}
J.~Solomon, K.~Crane, A.~Butscher, and C.~Wojtan, ``A general framework for
  bilateral and mean shift filtering,'' \emph{arXiv preprint arXiv:1405.4734},
  2014.

\bibitem{ZhangDZBL15}
W.~Zhang, B.~Deng, J.~Zhang, S.~Bouaziz, and L.~Liu, ``Guided mesh normal
  filtering,'' \emph{Comput. Graph. Forum}, vol.~34, no.~7, pp. 23--34, 2015.

\bibitem{WangFLTLG15}
P.~Wang, X.~Fu, Y.~Liu, X.~Tong, S.~Liu, and B.~Guo, ``Rolling guidance normal
  filter for geometric processing,'' \emph{{ACM} Trans. Graph.}, vol.~34,
  no.~6, p. 173, 2015.

\bibitem{ham2015}
B.~Ham, M.~Cho, and J.~Ponce, ``Robust image filtering using joint static and
  dynamic guidance,'' in \emph{{CVPR}}, 2015.

\bibitem{Taubin1995}
G.~Taubin, ``A signal processing approach to fair surface design,'' ser.
  SIGGRAPH '95, 1995, pp. 351--358.

\bibitem{Desbrun1999}
M.~Desbrun, M.~Meyer, P.~Schr\"{o}der, and A.~H. Barr, ``Implicit fairing of
  irregular meshes using diffusion and curvature flow,'' ser. SIGGRAPH '99,
  1999, pp. 317--324.

\bibitem{Taubin2001}
G.~Taubin, ``Linear anisotropic mesh filtering,'' IBM Research, IBM Research
  Report RC22213(W0110-051), 2001.

\bibitem{Ohtake2001}
Y.~Ohtake, A.~Belyaev, and I.~Bogaevski, ``Mesh regularization and adaptive
  smoothing,'' \emph{Computer-Aided Design}, vol.~33, no.~11, pp. 789--800,
  2001.

\bibitem{ChuangK11}
M.~Chuang and M.~M. Kazhdan, ``Interactive and anisotropic geometry processing
  using the screened poisson equation,'' \emph{{ACM} Trans. Graph.}, vol.~30,
  no.~4, pp. 57:1--57:10, 2011.

\bibitem{paris2009bilateral}
S.~Paris, P.~Kornprobst, J.~Tumblin, and F.~Durand, \emph{Bilateral filtering:
  Theory and applications}.\hskip 1em plus 0.5em minus 0.4em\relax Now
  Publishers Inc, 2009.

\bibitem{Jones2004}
T.~Jones, F.~Durand, and M.~Zwicker, ``Normal improvement for point
  rendering,'' \emph{IEEE Computer Graphics and Applications}, vol.~24, no.~4,
  pp. 53--56, 2004.

\bibitem{Wang2006}
C.~C. Wang, ``Bilateral recovering of sharp edges on feature-insensitive
  sampled meshes,'' \emph{IEEE Trans. Vis. Comput. Graphics}, vol.~12, no.~4,
  pp. 629--639, 2006.

\bibitem{KopfCLU07}
J.~Kopf, M.~F. Cohen, D.~Lischinski, and M.~Uyttendaele, ``Joint bilateral
  upsampling,'' \emph{{ACM} Trans. Graph.}, vol.~26, no.~3, p.~96, 2007.

\bibitem{xu2011image}
L.~Xu, C.~Lu, Y.~Xu, and J.~Jia, ``Image smoothing via {$L_0$} gradient
  minimization,'' \emph{ACM Trans. Graph.}, vol.~30, no.~6, pp. 174:1--174:12,
  2011.

\bibitem{RUDIN1992}
L.~I. Rudin, S.~Osher, and E.~Fatemi, ``Nonlinear total variation based noise
  removal algorithms,'' \emph{Phys. D}, vol.~60, no. 1-4, pp. 259--268, 1992.

\bibitem{Taubin12c}
G.~Taubin, ``Introduction to geometric processing through optimization,''
  \emph{{IEEE} Computer Graphics and Applications}, vol.~32, no.~4, pp. 88--94,
  2012.

\bibitem{he2013mesh}
L.~He and S.~Schaefer, ``Mesh denoising via {$L_0$} minimization,'' \emph{ACM
  Trans. Graph.}, vol.~32, no.~4, pp. 64:1--64:8, 2013.

\bibitem{Zhang2015Variational}
H.~Zhang, C.~Wu, J.~Zhang, and J.~Deng, ``Variational mesh denoising using
  total variation and piecewise constant function space,'' \emph{IEEE
  Transactions on Visualization and Computer Graphics}, vol.~21, no.~7, pp.
  873--886, 2015.

\bibitem{Perona90}
P.~Perona and J.~Malik, ``Scale-space and edge detection using anisotropic
  diffusion,'' \emph{IEEE Trans. Pattern Anal. Mach. Intell.}, vol.~12, no.~7,
  pp. 629--639, 1990.

\bibitem{sun2009concise}
J.~Sun, M.~Ovsjanikov, and L.~Guibas, ``A concise and provably informative
  multi-scale signature based on heat diffusion,'' in \emph{Proceedings of the
  Symposium on Geometry Processing}, 2009, pp. 1383--1392.

\bibitem{Vallet2008}
B.~Vallet and B.~L\'{e}vy, ``Spectral geometry processing with manifold
  harmonics,'' \emph{Computer Graphics Forum}, vol.~27, no.~2, pp. 251--260,
  2008.

\bibitem{zhang2010spectral}
H.~Zhang, O.~Van~Kaick, and R.~Dyer, ``Spectral mesh processing,''
  \emph{Computer Graphics Forum}, vol.~29, no.~6, pp. 1865--1894, 2010.

\bibitem{sun2007fast}
X.~Sun, P.~L. Rosin, R.~R. Martin, and F.~C. Langbein, ``Fast and effective
  feature-preserving mesh denoising,'' \emph{IEEE Trans. Vis. Comput.
  Graphics}, vol.~13, no.~5, pp. 925--938, 2007.

\bibitem{Pukelsheim1994}
F.~Pukelsheim, ``The three sigma rule,'' \emph{The American Statistician},
  vol.~48, no.~2, pp. 88--91, 1994.

\bibitem{Bouaziz2012}
S.~Bouaziz, M.~Deuss, Y.~Schwartzburg, T.~Weise, and M.~Pauly, ``Shape-up:
  Shaping discrete geometry with projections,'' \emph{Computer Graphics Forum},
  vol.~31, no.~5, pp. 1657--1667, 2012.

\bibitem{eigenweb}
G.~Guennebaud, B.~Jacob \emph{et~al.}, ``Eigen v3,''
  http://eigen.tuxfamily.org, 2010.

\bibitem{Sorkine2004}
O.~Sorkine, D.~Cohen-Or, Y.~Lipman, M.~Alexa, C.~R\"{o}ssl, and H.-P. Seidel,
  ``Laplacian surface editing,'' in \emph{Proceedings of the 2004
  Eurographics/ACM SIGGRAPH Symposium on Geometry Processing}, ser. SGP '04,
  2004, pp. 175--184.

\bibitem{Vasa2011}
L.~Vasa and V.~Skala, ``A perception correlated comparison method for dynamic
  meshes,'' \emph{IEEE Transactions on Visualization and Computer Graphics},
  vol.~17, no.~2, pp. 220--230, 2011.

\end{thebibliography}

\begin{IEEEbiography}[{\includegraphics[width=\columnwidth,clip]{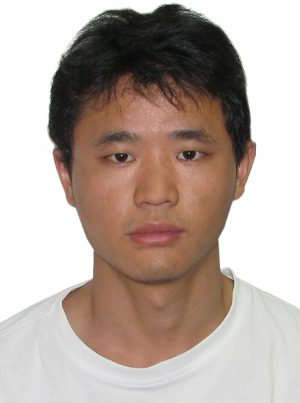}}]{Juyong Zhang}
	is an associate professor in the School of Mathematical Sciences at University of Science and Technology of China. He received the BS degree from University of Science and Technology of China in 2006, and the PhD degree from Nanyang Technological University, Singapore. His research interests include computer graphics, computer vision, and numerical optimization. He is a member of the IEEE.
\end{IEEEbiography}

\vspace*{-2em}

\begin{IEEEbiography}[{\includegraphics[width=\columnwidth,clip]{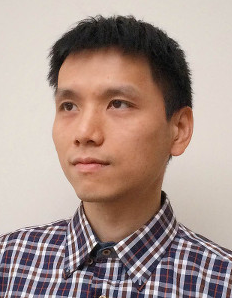}}]{Bailin Deng}
	is a lecturer in the School of Computer Science and Informatics at Cardiff University. He received the BEng degree in computer software (2005) and the MSc degree in computer science (2008) from Tsinghua University (China), and the PhD degree in technical mathematics from Vienna University of Technology (Austria). His research interests include geometry processing, numerical optimization, computational design, and digital fabrication. He is a member of the IEEE.
\end{IEEEbiography}

\vspace*{-2em}

\begin{IEEEbiography}[{\includegraphics[width=\columnwidth,clip]{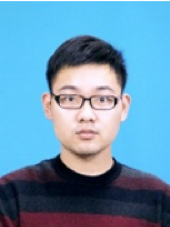}}]{Yang Hong}
	is currently working toward a master's degree in the School of Mathematical Sciences,
	University of Science and Technology of China. His research interests include geometric \& image processing and 3D modeling.
\end{IEEEbiography}

\vspace*{-2em}

\begin{IEEEbiography}[{\includegraphics[width=\columnwidth,clip]{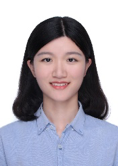}}]{Yue Peng}
	received her BS degree from the University of Science and Technology of China in 2016. Currently, she is working toward a master's degree in the USTC. Her research interests include computer graphics and geometry processing.
\end{IEEEbiography}

\vspace*{-2em}

\begin{IEEEbiography}[{\includegraphics[width=\columnwidth,clip]{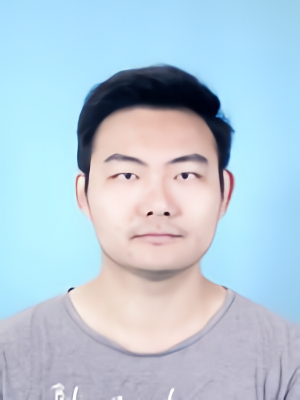}}]{Wenjie Qin}
	received the bachelor's degree from the University of Science and Technology of China in 2012. He is currently working toward a master degree in the University of Science and Technology of China. His research interests include geometric processing and filtering.
\end{IEEEbiography}

\vspace*{-2em}

\begin{IEEEbiography}[{\includegraphics[width=\columnwidth,clip]{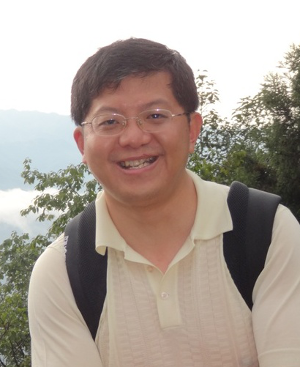}}]{Ligang Liu}
	is a Professor at the School of Mathematical Sciences, University of Science and Technology of China. His research interests include digital geometric processing, computer graphics, and image processing. He serves as the associated editors for journals of IEEE Transactions on Visualization and Computer Graphics, IEEE Computer Graphics and Applications, Computer Graphics Forum, Computer Aided Geometric Design, and The Visual Computer. He served as the conference co-chair of GMP 2017 and the program co-chairs of GMP 2018, CAD/Graphics 2017, CVM 2016, SGP 2015, and SPM 2014. His research works could be found at his research website: {http://staff.ustc.edu.cn/∼lgliu}.
\end{IEEEbiography}

\end{document}